\documentclass[10pt,journal,compsoc]{IEEEtran}
\ifCLASSOPTIONcompsoc
  \usepackage[nocompress]{cite}
  \usepackage{comment}
  \usepackage[hyphens]{url}
  \usepackage{amsmath,amssymb,amsfonts}
  \usepackage{algorithmic}
  \usepackage{graphicx}
  \usepackage{textcomp}
  \usepackage{xcolor}
  \usepackage{comment}
  \usepackage{array}
  \usepackage{multirow}
  \usepackage{caption}
  \usepackage{subcaption}
  \usepackage{tikz}
  \usepackage{cite}
  \usepackage{adjustbox}
\usepackage{soul}
\usepackage{amsmath,amssymb,amsfonts}
\usepackage{booktabs}
\usepackage{tabularx}
\usepackage{bbding}
\usepackage{pifont}
\usepackage{wasysym}
\usepackage{amssymb}
\def\BibTeX{{\rm B\kern-.05em{\sc i\kern-.025em b}\kern-.08em
    T\kern-.1667em\lower.7ex\hbox{E}\kern-.125emX}}
\definecolor{green}{rgb}{0.09, 0.45, 0.27}
\usepackage{comment}
\usepackage{setspace}
\usepackage{todonotes}
\usepackage{pdfcomment}
\usepackage{fancyhdr}
\newcommand{\vittorio}[2][]{\todo[color=red!50, , inline, #1]{\textbf{Vittorio:} #2}}

\newlength\MAX  

\newcommand{\DrawPercentageBar}[1]{%
  \begin{tikzpicture}
    \fill[color=black]   (0.0 , 0.0) rectangle (#1*3ex , 1.5ex );
    \fill[color=gray] (#1*3ex  , 0.0) rectangle (3.0ex, 1.5ex);
  \end{tikzpicture}%
}

\newcommand*\circled[1]{\tikz[baseline=(char.base)]{
            \node[shape=circle,draw,inner sep=1pt,font=\sffamily\footnotesize] (char) {\textbf{#1}};}}

\usepackage[most]{tcolorbox}

\usepackage[all=normal,floats,leading,charwidths,bibbreaks,mathspacing]{savetrees}
\else
  \usepackage{cite}
  \usepackage{comment}
\fi
\ifCLASSINFOpdf
\else
\fi
\begin{document}
%

\title{Laccolith: Hypervisor-Based\\Adversary Emulation with Anti-Detection}
%
%
%
%

\author{Vittorio~Orbinato, Marco Carlo Feliciano, Domenico Cotroneo, Roberto Natella
\IEEEcompsocitemizethanks{\IEEEcompsocthanksitem V. Orbinato, D. Cotroneo, and R. Natella are with the Department of Electrical Engineering and Information Technology (DIETI), Università degli Studi di Napoli Federico II, Naples, Italy.\protect\\
E-mail: \{vittorio.orbinato, cotroneo, roberto.natella\}@unina.it
\IEEEcompsocthanksitem M.C. Feliciano is with Secureware s.r.l., Naples, Italy.\protect\\
E-mail: mfeliciano@secware.it}
}

%
%


\markboth{IEEE Transactions on Dependable and Secure Computing, DOI: \href{https://doi.org/10.1109/TDSC.2024.3376129}{10.1109/TDSC.2024.3376129}}{}%
\IEEEtitleabstractindextext{%
\begin{abstract}
Advanced Persistent Threats (APTs) represent the most threatening form of attack nowadays since they can stay undetected for a long time. Adversary emulation is a proactive approach for preparing against these attacks. However, adversary emulation tools lack the anti-detection abilities of APTs. 
We introduce Laccolith, a hypervisor-based solution for adversary emulation with anti-detection to fill this gap. We also present an experimental study to compare Laccolith with MITRE CALDERA, a state-of-the-art solution for adversary emulation, against five popular anti-virus products. We found that CALDERA cannot evade detection, limiting the realism of emulated attacks, even when combined with a state-of-the-art anti-detection framework. Our experiments show that Laccolith can hide its activities from all the tested anti-virus products, thus making it suitable for realistic emulations.
\end{abstract}

}

\maketitle


\IEEEdisplaynontitleabstractindextext

%
\IEEEpeerreviewmaketitle

\IEEEraisesectionheading{\section{Introduction}}

\label{sec:introduction}
\emph{Advanced Persistent Threats} (APTs) have become a severe threat in several domains where the impact can be exceedingly high (e.g., in terms of service outages, private data breaches, and intellectual property theft), such as healthcare, manufacturing, telecom, energy, and transportation. In APT, attackers accomplish their goals through a carefully planned sequence of malicious actions performed while being careful to stay undetected. These attacks are getting even more challenging to counter, as they are carried out by cybercriminal and state-sponsored groups. 
Well-known examples include the Stuxnet attack, which has been sabotaging Iran's nuclear centrifuges since 2005 and was uncovered in 2010 \cite{StuxnetDossier}, GhostNet \cite{GhostNet}, and Carbanak \cite{Carbanak}. The time an APT attack goes undetected (``dwell time'') has been estimated to be up to 700 days \cite{MandiantTrends}. With all this time to act, APTs can cause irreparable damage.


\emph{Adversary emulation} is the most effective prevention against APTs \cite{Applebaum2016IntelligentAR}. Adversary emulation is a proactive approach that reproduces the actions of an APT inside a target computer infrastructure. It grants several advantages for security training and assessment purposes, such as understanding threats, realistic and comprehensive testing, identification of weaknesses, and validation of security controls. This approach focuses on post-compromise scenarios, assuming that the APT has already gained access to the target infrastructure and is infecting hosts and devices silently, escalating privileges, and exfiltrating data. Adversary emulation is usually exercised in virtualized environments (\emph{cyber-ranges}) and, in general, it encompasses time-consuming activities that engage human personnel (\emph{red teams}). 
Several tools aim to automate adversary emulation \cite{zilberman2020sok}, which offer automated procedures that implement APT techniques, as learned from threat intelligence sources. These tools typically automate actions for information gathering, lateral movement across hosts, connections to command-and-control servers, and privilege escalation. 

A fundamental characteristic of APTs is the adoption of \emph{anti-detection} techniques, to hide their traces from AV (anti-virus) products and EDR (endpoint detection and response), a modern evolution of AVs, and to persist in the target infrastructure as long as possible. 
Therefore, adversary emulation also needs to apply anti-detection techniques, in order to perform realistic and fruitful security assessments and training activities. Examples of anti-detection techniques include un-hooking probes used by EDRs to instrument and monitor DLL and API uses \cite{HookingMemoryProtection}; disabling or hampering event tracers, such as the Event Tracing for Windows (ETW) subsystem \cite{ETWbypass}; using malicious kernel modules to hide processes and files \cite{Outflank}; obfuscating malicious payloads (e.g., shellcodes) \cite{EvadingShellcode, EDRShellcode}. 
However, it is challenging for adversary emulation tools to automate anti-detection techniques, since both APTs and AV solutions are continuously evolving, in a ``cat-and-mouse'' game. 
Currently, adversary emulation tools have to be customized for the specific AV/EDR to evade, which requires considerable skills and development efforts \cite{iRedTeam, Cyberstruggle}, and is prone to become outdated and ineffective. For these reasons, emulating anti-detection techniques in automated ways in adversary simulations is still impractical. This is a significant limitation in realistically emulating APTs.

In this work, we investigate a novel solution for adversary emulation with anti-detection capabilities, to avoid the previously mentioned ``cat-and-mouse'' game. We tested state-of-the-art solutions for adversary emulation (\emph{MITRE CALDERA} \cite{zilberman2020sok} 
, \emph{Atomic Red Team} \cite{AtomicRedTeam}, \emph{Invoke-Adversary} \cite{InvokeAdversary}
) and for anti-detection (\emph{Inceptor} \cite{Inceptor}) against multiple AV/EDR products. We found that several malicious actions (and even the installation of the emulation agent) cannot evade detection, thus limiting the realism of emulated attacks. Laccolith is based on a novel \emph{hypervisor-based} architecture for adversary emulation. This design choice was led by the fact that cybersecurity exercises typically happen in virtualized environments \cite{yamin2020cyber, beuran2018integrated, vceleda2015kypo, wroclawski2016deterlab, ferguson2014national}. 

Our solution (\emph{Laccolith}) enables the non-detectable execution of malicious actions, by injecting them from the lowest layers of the software stack. We experimented with Laccolith against several AV solutions for Microsoft Windows. The results showed that Laccolith was able to execute all of the malicious actions, which evaded all of the tested AV products. Our solution does not require customizations for the specific version of the target system, as it can reliably execute non-detectable actions across different versions of the guest OS and AV products. 

In summary, the research contributions of this work are:
\begin{itemize}
    \item An innovative architecture for adversary emulation with anti-detection capabilities;
    \item A working prototype of the proposed architecture;
    \item An experimental analysis of anti-detection capabilities in state-of-the-art solutions for adversary emulation;
    \item An evaluation of the proposed solution compared to the state-of-the-art.
\end{itemize}

The remainder of the paper is structured as follows. In Section \ref{sec:background}, we elaborate on background concepts about APTs and adversary emulation. In Section \ref{sec:solution}, we present the proposed solution for adversary emulation. In Section \ref{sec:experiment}, we illustrate experiments on the anti-detection of existing solutions and our proposed one. In Section \ref{sec:related}, we discuss related work. Section \ref{sec:conclusion} concludes the paper.

\section{Background}
\label{sec:background}

APTs are complex attack campaigns that eventually gain access to a target infrastructure due to inevitable weaknesses, such as the exploitation of an unknown (``zero-day'') software vulnerability, a known vulnerability in outdated software, or human mistakes that result in weak credentials, information disclosure, and execution of untrusted code \cite{MITREAnalytics}. Initial access represents just the preliminary part of the attack, followed by several other activities that depend on the objectives and skills of the attackers. 

In recent years, several models were introduced to describe the actions and behavior of attackers. 
ATT\&CK \cite{attack} is a security framework introduced by the MITRE Corporation and has rapidly become the \textit{de facto} standard in the security community. ATT\&CK provides a detailed classification of the adversaries' tactics, techniques, and procedures (TTPs), organized as a matrix, continuously updated by observing and analyzing threat intelligence sources \cite{strom2018mitre, zilberman2020sok}. \emph{Tactics} describe \textbf{why} an adversary performs an action, and \emph{techniques} describe \textbf{how} they do it. Techniques are described in the ATT\&CK model from both offensive and defensive points of view, so they are a useful reference and pivot between both disciplines. The descriptions also include references to known usage examples of the specific techniques and links to public threat reporting on adversary groups and their campaigns. Moreover, techniques can further specialize into sub-techniques, which usually provide more specific versions of an attacker's actions. Finally, \emph{procedures} represent the actual implementation of techniques for specific target systems.
At the time of writing, ATT\&CK encompasses 193 techniques and 401 sub-techniques. The richness of this framework makes it suitable for modeling APT campaigns. It is worth noting that a great number of combinations of TTPs is possible, making it relevant to focus on the TTPs of specific APTs of interest. Typically, APTs are classified according to the targeted domain, such as education, finance, government, and healthcare.

\begin{figure}[htbp]
\centering
\centerline{\includegraphics[width=\linewidth]{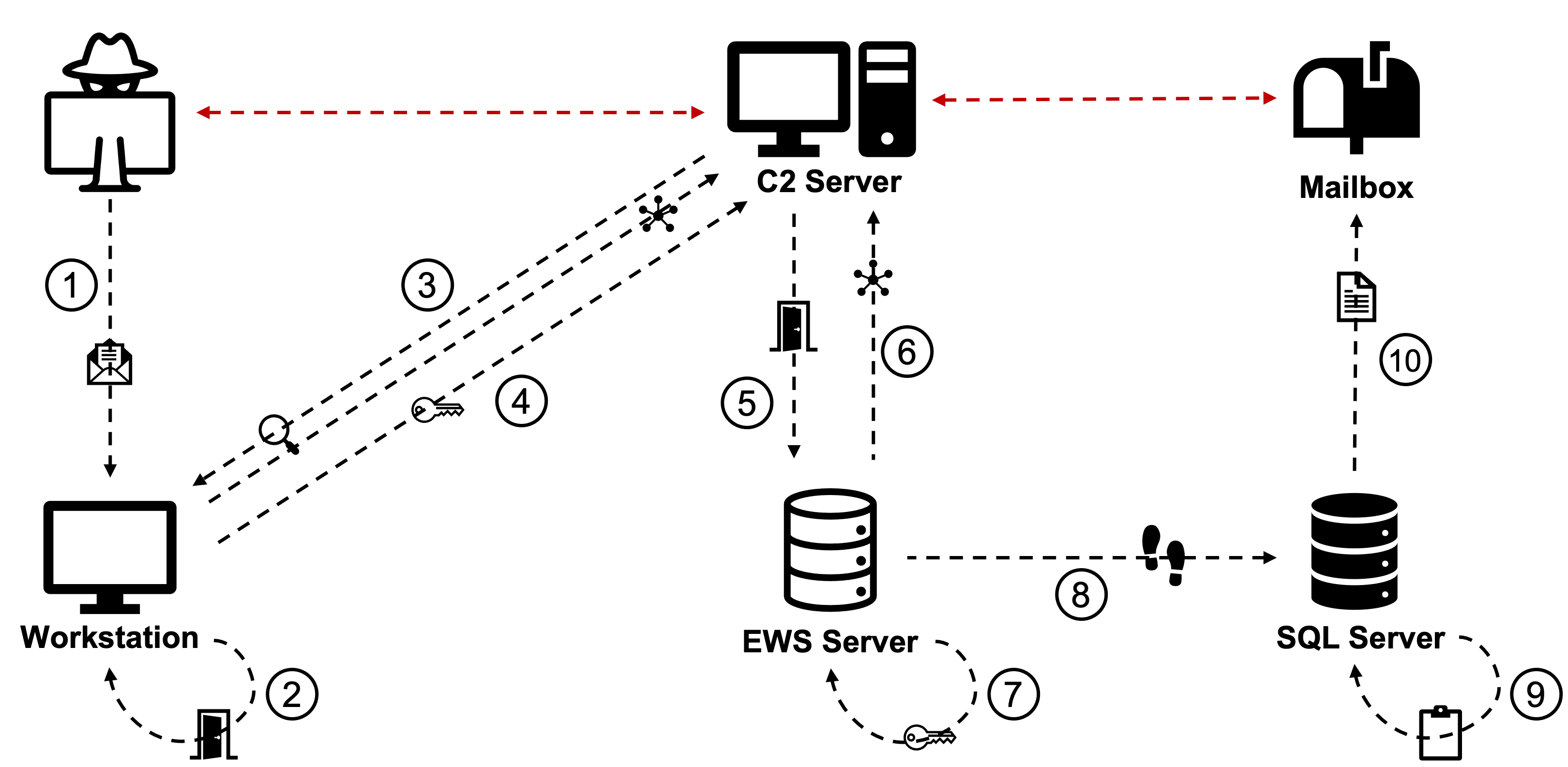}}
\caption{OilRig Operational Steps. The \textcolor{red}{red dotted} arrows represent C2 communication, the black ones the operational steps.}
\label{fig:OilRig}
\end{figure}

As an example, we will refer to the OilRig APT \cite{oilrig}. OilRig is a cyber threat actor operating since 2014, interested in several domains, such as finance, government, energy, chemical, and telecommunications \cite{oilrigsummary}. It primarily leverages social engineering as an initial attack vector. The Center for Threat-Informed Defense (CTID) \cite{CTID} defined the model of its behavior and actions by collecting and analyzing threat and incident reports from several sources, such as Cyware \cite{Cyware}, Mandiant \cite{OilRigMandiant}, and Malwarebytes Labs \cite{Malwarebytes}. Figure \ref{fig:OilRig} shows OilRig's operational steps for \emph{exfiltrating data from a targeted server}. The first tactical goal is \emph{Initial Access}, to identify the target to exploit: OilRig achieves it using \emph{spearfishing attachment} (T1566\footnote{\textbf{Txxxx} indicates a technique from the MITRE ATT\&CK framework} \cite{T1566}) (step \circled{1} in Figure \ref{fig:OilRig}). Once the target opens the malicious attachment (T1204.002 \cite{T1204}), a backdoor is installed on the host machine (step \circled{2}). Then, it performs \emph{enumeration} (T1087 \cite{T1087}) (step \circled{3}), leading to the discovery that the user is a member of the administrator group on an Exchange Web Server (EWS). Using the aforementioned backdoor, OilRig \emph{obtains credentials} to EWS (T1003 \cite{T1003}) (step \circled{4}). Leveraging these credentials, the attackers connect to EWS and install a new backdoor (step \circled{5}) to perform further \emph{enumeration} (T1087 \cite{T1087}) (step \circled{6}): this leads to the discovery of an SQL server. Using the backdoor, OilRig \emph{dumps the credentials} (T1003 \cite{T1003}) (step \circled{7}) to access the SQL server. The next step is \emph{lateral movement} towards the server (step \circled{8}) by passing-the-hash (T1550.002 \cite{T1550}), where the attackers will copy the database backup files (step \circled{9}) and exfiltrate them to a controlled mailbox (T1048 \cite{T1048}) (step \circled{10}).

Once a model of an APT has been created using the ATT\&CK framework, it can be used for testing and assessing the security posture of a target infrastructure through adversary emulation. Several tools have been developed to facilitate and automate adversary emulation. They provide automated low-level procedures that implement the techniques described by the ATT\&CK framework. These procedures emulate the actions of actual adversaries in a typical attack scenario. In particular, we will refer to CALDERA \cite{Applebaum2016IntelligentAR, CALDERAGitHub}, an open-source adversary emulation tool developed by MITRE. CALDERA has gained popularity due to its feature-richness, maturity, and ease of setup and use \cite{zilberman2020sok}. In addition, it provides high coverage of TTPs of the MITRE ATT\&CK framework. In contrast, other adversary emulation tools consist of a small set of scripts (e.g., in Python and Powershell) that automate individual malicious actions (e.g., the Atomic Red Team), leaving human red teams to string these single actions into a complete APT campaign.

\begin{figure}[htbp]
\centering
\centerline{\includegraphics[width=\linewidth]{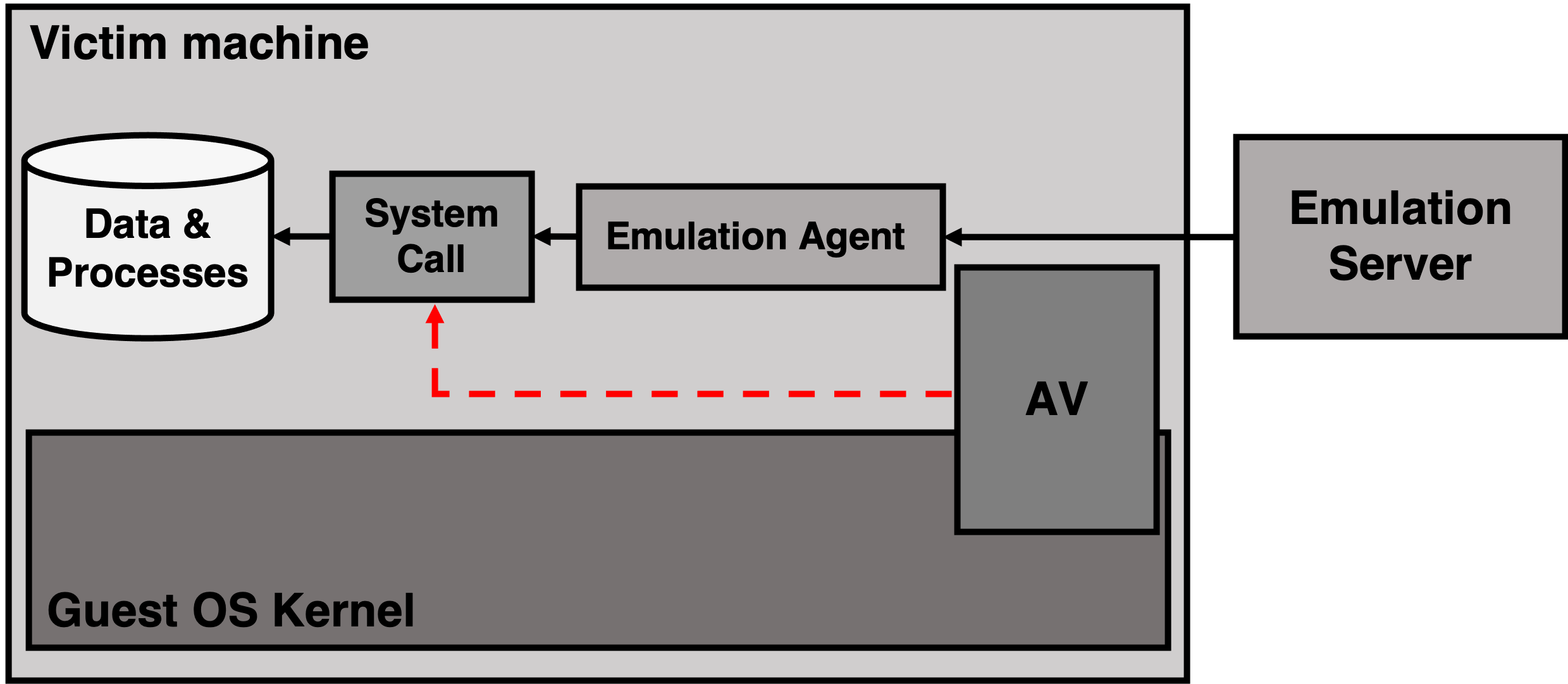}}
\caption{Adversary emulation traditional architecture.}
\label{fig:Adversary Emulation architecture}
\end{figure}

CALDERA is based on a client/server architecture, as shown in Figure \ref{fig:Adversary Emulation architecture}: the command-and-control (C2) server is responsible for administrating the operations, i.e., the simulations, planning the actions to execute, and making all the information available to the user. On the other hand, the clients are the agents installed on the target machines: the agent is responsible for the actual execution of the actions and for sending the results back to the C2 server. 

Adversary emulation tools, such as CALDERA, do not provide anti-detection techniques, 
hence the need to turn off the AVs during the emulations.
Turning off the AV causes the system to run in a different environment. This aspect constitutes a relevant concern in critical domains (such as defense), which need to accurately emulate advanced adversaries with evasion abilities. For instance, for forensic analysis purposes, it is best not to alter the lists of processes and files seen by the security teams, which would be polluted by adversary emulation tools that run in user-space. Moreover, the emulation designers may want some specific malicious activities (not all) to be detected, to guide the participants in solving pre-planned cyber-attack scenarios. As an extreme, if none of the activities were detectable, the blue teams would not have any element to learn from. Our solution allows the designer to selectively configure which activities to hide from the AVs and the participants.

To perform non-detectable actions, red teams hide their traces using separate tools. A state-of-the-art toolkit for anti-detection is Inceptor \cite{Inceptor}. Inceptor is a well-known template-driven framework for AV evasion. A template is a generic, customizable loader with placeholders for evasion techniques and the actual payloads. There are many templates for three different types of payloads: .NET, PowerShell, and native code. Inceptor offers the possibility of chaining encoding techniques to evade static code analysis. It is also possible to plugin additional source code writing techniques to evade the Windows Antimalware Scanning Interface's (AMSI) dynamic analysis \cite{hendler2020amsi}. AMSI also analyses in-memory artifacts, for example, text areas the code is going to jump to. For this reason, Inceptor includes AMSI bypass techniques \cite{40}. 
In our experimental evaluation of the state-of-the-art on adversary emulation and anti-detection, we consider MITRE CALDERA in combination with Inceptor.
\section{Proposed solution}
\label{sec:solution}
In this section, we present our proposed solution for adversary emulation with anti-detection. We first provide an overview, including our system model and assumptions for our design. We then describe in detail the proposed design. Finally, we provide information about our implementation.

\subsection{Overview}
\label{sec:overview}
We based the design of Laccolith on a set of assumptions applied in general for adversary emulation. The assumptions include:

\begin{itemize}
    \item Adversary emulation focuses on post-compromise scenarios where the attacker has already gained a foothold inside the system (e.g., through phishing or exploiting a vulnerability) and is performing more malicious actions, such as gaining more privileges and stealing information.
    \item Adversary emulation is performed in the context of cybersecurity exercises, which are authorized and overseen by a ``white team'' in the organization (e.g., system administrators). Adversary emulation can be warranted system privileges as needed.
    \item Laccolith is designed to perform adversary emulation in a virtualized environment, where hosts are deployed using virtual machines. This is typically the case in cybersecurity exercises, where dedicated networks and hosts are deployed using virtualization infrastructures \cite{beuran2018integrated, schoonover2018galaxy, kouril2014cloud}. For example, all the most popular Cyber Ranges run in the cloud, leveraging virtualization technologies \cite{yamin2020cyber, ferguson2014national, vceleda2015kypo}. Moreover, the solution applies to organizations that run their own private data centers, which is often the case for high-criticality domains such as defense, critical infrastructures, and healthcare \cite{dhaya2021dynamic, regola2013storing}. Instead, the proposed solution is not meant for networks and services running on bare-metal hardware.
    \item Endpoints in the environment can be equipped with AV products, as in the case of real computer networks. This assumption is not met by traditional adversary emulation tools, which require turning off AV products to install and run the tools \cite{zilberman2020sok}. Laccolith is designed to overcome this limitation.
\end{itemize}

The driving idea for our design is to leverage kernel- and hypervisor-level privileges to execute malicious actions, which would otherwise be detected if performed from user level (i.e., from an application process).\footnote{The name \emph{Laccolith} reflects the injection of malicious actions from the lower layers of the software stack. A laccolith is a volcanic phenomenon where magma rises through the Earth's crust, forcing rock strata upward.}
AV solutions typically use ``hook'' functions to intercept invocations of the system calls. When applications (including malicious ones) invoke system calls, the hook functions are executed instead (e.g., by replacing pointers in the system call table), which can check the invocation and detect suspicious activities. In traditional adversary emulation (Figure \ref{fig:Adversary Emulation architecture}), an agent process executes malicious actions on behalf of a red team by issuing system calls that access OS resources, such as files, processes, connections, and others.
Without any anti-detection technique, AV products can detect these actions by checking system calls.

To perform non-detectable actions, it is necessary to use system calls not monitored by the AVs, hence from the kernel level. Consequently, the hypervisor level is the ideal choice to accomplish that, leveraging virtual machine introspection (VMI) techniques. In our design, we circumvent the anti-virus checks by introducing an emulation agent from the hypervisor. Since the hypervisor has full privileges on a physical machine, it can warrant full read and write access to the state of a virtual machine. We use these privileges to install an emulator into the kernel of the guest OS of the virtual machine. This way, it is possible to directly access guest memory, without issuing user-space system calls. From the agent installed in the guest OS kernel, we can perform malicious actions by calling kernel-level APIs. Such calls cannot be detected by AV products since they are not subject to security checks.

Moreover, we designed the kernel-level agent with the ability to run user-level commands, as in traditional adversary emulation tools. Therefore, red teams can combine kernel-level and user-level actions to perform both detectable and non-detectable actions. This flexibility enables red teams to perform realistic training exercises for security teams, where the emulated APT leaves only a few selected traces of malicious activity.

\begin{figure}[!htbp]
\centering
\centerline{\includegraphics[width=\linewidth]{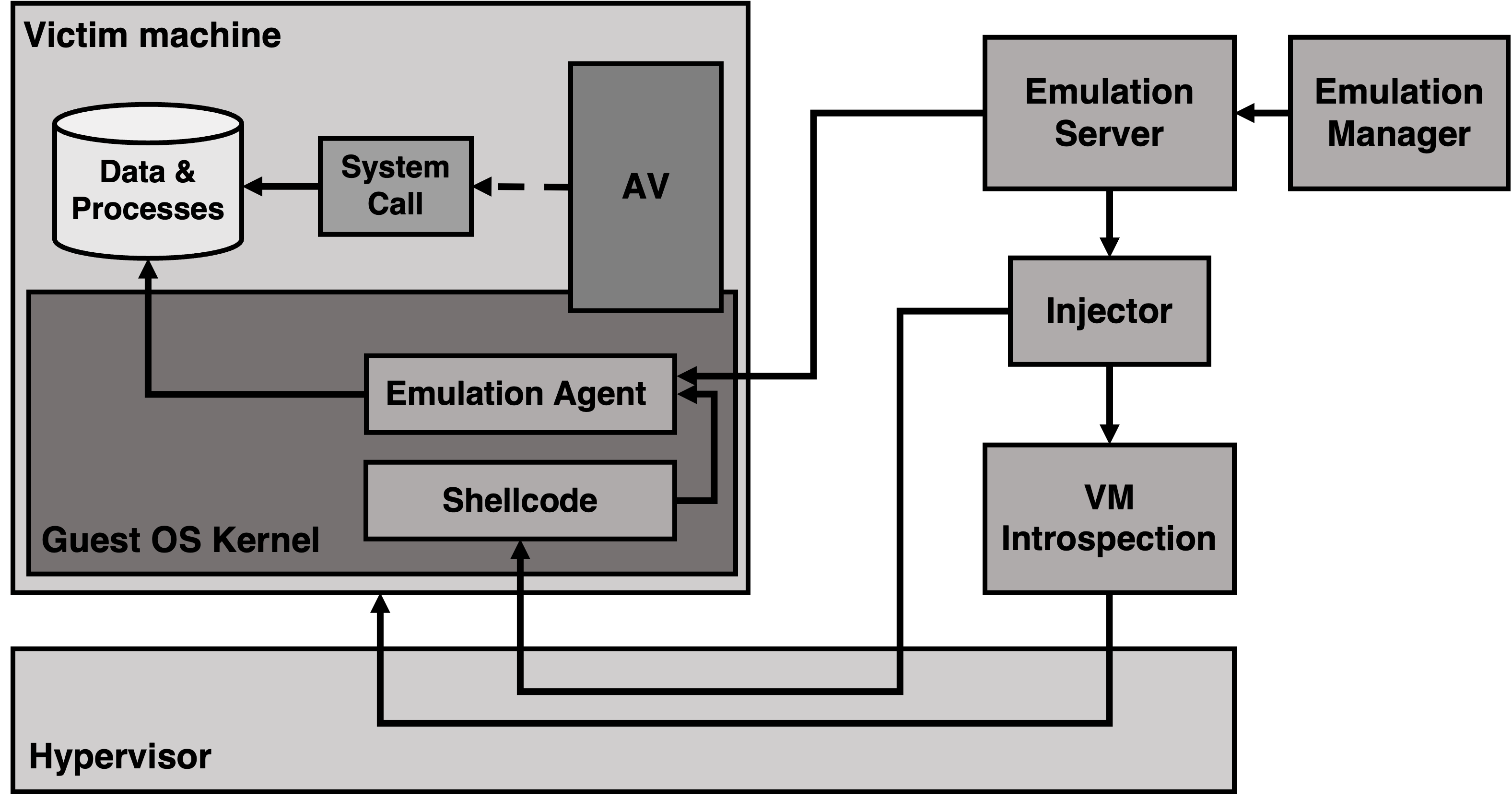}}
\caption{Laccolith architecture.}
\label{fig:Laccolith architecture}
\end{figure}


Laccolith offers a complete architecture for adversary emulation: a central \emph{emulation manager} orchestrates \emph{emulation agents} across the network and issues actions to execute on the victim hosts. Therefore, the red team controls the emulation agents to reproduce the actions of an APT campaign. The red team can interact with the emulation manager through web and command-line interfaces. 
On each physical machine, Laccolith installs an \emph{emulation server}, which runs at hypervisor-level, and manages the emulation agents on each physical machine. The emulation server installs the emulation agents on each virtual machine by leveraging VM introspection techniques to modify guest memory and create a new execution flow inside the guest kernel. 
Moreover, the emulation server forwards communication between the emulation agents and the manager. The communication traffic between the components of Laccolith is invisible to the participants of the cybersecurity exercise since this traffic flows at the physical level, beyond the virtual networks of the virtual machines.

Laccolith is neutral concerning post-facto analysis: the designer of the attack scenario can decide which artifacts should be left inside the system for forensics purposes, thanks to the possibility of configuring which specific actions are executed (un)detectably. For instance, the designer may want to stealthily poison some entries in the system registry for forensic analysis. Moreover, the designer can launch processes or connections in a non-detectable fashion from kernel space. In these cases, the participant is challenged to use event correlation techniques and SIEM technology to analyze the system state and understand what is happening.

In summary, our design addresses the following technical challenges: \emph{1)} to run an emulation agent in a victim machine, with no intrusive modifications of the system under analysis; \emph{2)} to perform adversarial actions without triggering AVs; and \emph{3)} to coordinate such adversary emulation campaigns from outside of the victim machine through a C2 infrastructure.

\subsection{Emulation server}
\label{sec:emulation_server}
The \textit{emulation server} is the trickiest component in our architecture. It is responsible for injecting \textit{emulation agents} inside the VMs, by leveraging read/write (R/W) access permissions to modify their state. It relies on installing a kernel-level agent to avoid detection from AV products.



\subsubsection{Overview of the injection method}
\label{sec:injection process}

The emulation agent needs to run in a dedicated area of the virtual memory of the VM. However, the allocation of virtual memory cannot be directly performed from the hypervisor but needs to be performed by the guest OS kernel, which is aware of the current layout of memory allocations. Consequently, the injection of the emulation agent consists of two stages. 
The first stage consists of the injection of a small program (a ``shellcode'') in kernel space, responsible for allocating the code region for the emulation agent, while the second stage entails bootstrapping the asynchronous execution of the agent in that area. The injector overwrites a piece of existing kernel code with the shellcode to make it executed by the kernel. Differently from the \emph{injector} (which runs in the hypervisor and can only work with ``guest physical'' memory addresses), the shellcode can allocate and access the ``guest virtual'' memory addresses since it runs within the guest OS kernel.
Before selecting the target kernel code to overwrite, we need to address two requirements for the shellcode:
\begin{enumerate}
    \item \textit{Space}: the shellcode should be small enough to fit into the target kernel code;
    \item \textit{Behavior}: if the overwritten kernel code is invoked, the shellcode needs to handle the call without hanging the calling process. Moreover, it should also handle the case of concurrent calls from different processes.
\end{enumerate}

We consider the code of system calls for the injection of the shellcode since their position in the virtual memory of the VM can be identified with \emph{virtual machine introspection} (VMI) techniques. VMI is an approach to gain visibility and control over VMs without modifying the guest OS \cite{garfinkel2003virtual}. 
Moreover, system calls are regularly invoked by applications, thus assuring that the shellcode will be eventually executed. 
The previously mentioned requirements are the reason why we cannot trivially overwrite a system call with the code of the \emph{emulation agent}: the first one would restrict the size of the agent, which would hamper the implementation of malicious actions; the second requirement implies that the system call code needs to be restored at some point, thus removing the injected code. 

To choose a target system call to overwrite, we look for a \textit{linear region of code}, i.e., a memory region that meets the following requirements:
\begin{enumerate}
    \item The memory region is from the code area of the guest OS kernel; this code is only executed by starting from the initial address of the memory region. For example, this requirement is satisfied if the memory region exactly matches the code of an individual kernel function; in this case, other kernel code only jumps to the initial address of the memory region (i.e., there are no jumps to addresses in the middle of the memory region).
    \item The code in the memory region does not call other functions. This requirement avoids that code not belonging to the memory region returns in the middle of the region.
    \item The region fits inside a memory page without crossing page boundaries.
\end{enumerate}

The first and second requirements allow us to inject arbitrary code in the memory region without risking the kernel jumping in the middle of the region, which would likely raise CPU exceptions (e.g., executing an invalid opcode). The shellcode will execute when the kernel executes the memory region, in place of the original code. 
The third requirement is necessary to make it easier to modify the VM memory from the hypervisor since code over multiple pages in virtual memory does not necessarily map to contiguous pages in physical memory. 
An example of an eligible memory region in Microsoft Windows OS is the \textit{MmQueryVirtualMemory} function (about $3,800$ bytes), which is called by the \textit{NtQueryVirtualMemory} system call. In general, linear regions of code are plentiful and easy to find: given the function call graph of the kernel code (e.g., \textit{ntoskrnl.exe} in Microsoft Windows), a linear region of code is a function without fan-out that fits within a single memory page. 

\subsubsection{Injection method}
\label{sec:detailed injection}

\begin{figure*}[!htbp]
\centering
\centerline{\includegraphics[width=\textwidth]{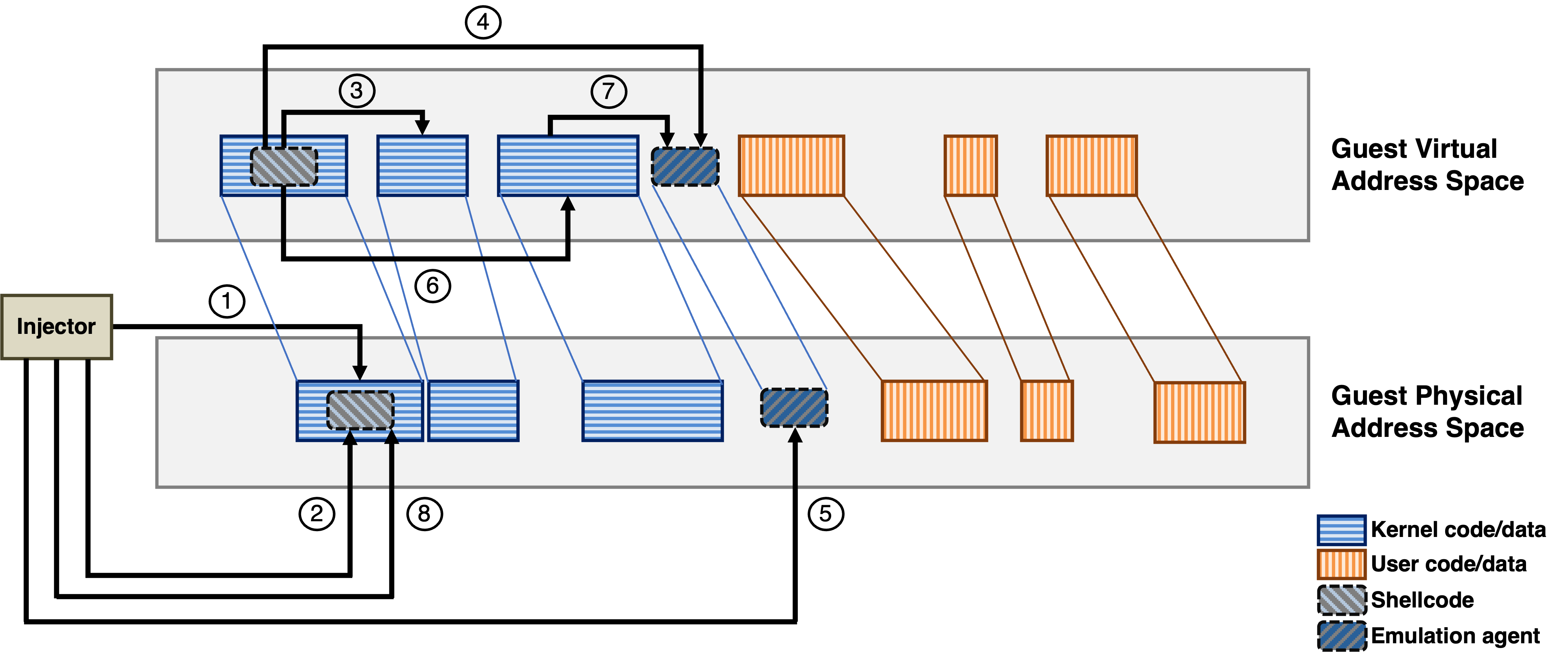}}
\caption{Injection method in Laccolith.}
\label{fig:Injection Chain}
\end{figure*}

The injection method is shown in Figure \ref{fig:Injection Chain} and summarized as follows. First, the \emph{injector} searches the code of the target system call in memory (step \circled{1}). Since the \emph{injector} can only access VM memory through guest physical addresses, it scans memory by looking for code that matches the initial unique bytes of the target system call. The injector leverages VMI to make this process more portable and efficient: it determines from VMI the unique bytes of the system call from the kernel binary, and the relative offset of the system call within the kernel virtual address space. Then, it only scans memory pages on that relative offset. 
Once the position of the target system call has been found, the \emph{injector} overwrites it with the shellcode (step \circled{2}). 

Then, the shellcode will eventually execute when the system call is invoked. 
The shellcode will execute for a limited time, long enough to allocate a memory area and communicate to the emulation server that the memory area is ready for loading the emulation agent (the last step of the method). 
The shellcode uses the APIs of the guest OS kernel (step \circled{3}) to allocate a contiguous area of guest virtual memory (for example, the \textit{MmAllocateContiguousMemory} function from the Windows Driver Kit (WDK) \cite{22}). 
The shellcode then obtains pages of contiguous virtual memory (e.g., $16$KB in our implementation) from the guest virtual address space (step \circled{4}), without the constraints that previously applied for the shellcode. 

During this time window, any other concurrent call to the target system call must immediately return, since the shellcode should execute only once. For this purpose, the shellcode acquires a spinlock in a \emph{non-blocking} way, without busy-waiting if the spinlock is already held (e.g., using the \textit{xchg} instruction on Intel architectures \cite{31}). Once acquired, the shellcode holds the spinlock to avoid concurrent executions.

Next, the shellcode writes a pre-defined value (``egg''), also known by the injector, within the allocated memory area. This way, the injector can find the allocated memory area using only guest physical addresses. 
The injector overwrites the allocated memory with the code of the emulation agent (step \circled{5}). Finally, the shellcode sets up an execution context to run the emulation agent in the kernel. Again, the shellcode uses APIs of the guest OS kernel to create a kernel thread (step \circled{6}), such as using the \textit{PsCreateSystemThread} function in Windows. The new kernel thread will be configured to run the code of the emulation agent in the allocated memory area (step \circled{7}), e.g., using the \textit{StartRoutine} of the Windows kernel. As a result, the emulation agent will execute with high privileges since it will run in kernel mode. After the agent starts, the injector restores the original system call code to leave no traces in the target VM other than the emulation agent (step \circled{8}).

\subsection{Emulation agent}
\label{sec:emulation_agent}

The \textit{emulation agent} is responsible for receiving and executing commands from the emulation server. Since it runs with high privileges, it can call kernel APIs as if it were a kernel module. The agent can also access the OS resources, such as the process descriptors and the system registry. Through these resources, the agent can access the memory of user-level processes since a process descriptor indicates the memory regions where the code and data of a process are. It can both read (e.g., process dumping to steal passwords, tokens, and other data) or write (e.g., process/DLL injection to hide malicious code in system processes) the processes' memory, and modify the system configuration, e.g., changing the registry for persistence purposes. Moreover, the agent can read/write the file system, e.g., navigating through folders, creating new files, and deleting existing ones. It can also create new user-level processes to execute commands, such as simulating fileless malware by running system binaries (e.g., PowerShell, WMI) and creating new OS resources. For example, the execution of user-mode commands can be accomplished using worker factories \cite{WorkerFactoriesBlog, WorkerFactoriesPoC}, a Windows mechanism that allows the kernel to create user-space thread pools and let them execute specific tasks.



\subsection{Emulation manager}
\label{sec:emulation_manager}
The \textit{emulation manager} communicates with the agents. Its role is critical to orchestrating the low-level actions performed by the agents into emulating the complex behavior of APTs, as in the example of Section \ref{sec:background}. 
It also offers a user interface to manage Laccolith. In particular, it offers interfaces to inject agents into specific targets, to access C2 functions (e.g., listing connections to agents, sending a command to an agent, reading its output, starting an autonomous operation), to customize parameters (e.g., choosing which payload to inject), and upload and download files (e.g., to exfiltrate data and information gathering). The emulation manager is also responsible for the management of the \textit{facts}. A fact is a piece of information about the target system, helpful to run an ability. The manager communicates the fact values to emulation agents when they need specific information to perform an action. For example, suppose the goal of the emulation is to perform lateral movement to another machine in the local network. The emulation agent will perform a network scanning action, whose outcome will be a fact containing the username and IP address of the target machine. The agent will then use this fact to perform lateral movement.


\subsection{Implementation}
\label{sec:implementation}

We implemented the design of Laccolith for the QEMU hypervisor for x86\_64 \cite{QEMU}, running on a Linux host managed by Libvirt \cite{LibVirt}, and with hardware-supported virtualization based on KVM \cite{KVM}.\footnote{
The source code and documentation of the prototype are available at \url{https://github.com/dessertlab/Laccolith} for research purposes. A commercial version of the tool and support is provided by the Secureware s.r.l. spin-off company, see \url{https://www.secware.it/}.}
Laccolith runs as a privileged process on the Linux host and accesses VM memory through a virtual device file. Our implementation targets the Microsoft Windows OS as the guest OS for the victim machine. 
We used Volatility \cite{Volatility}, a virtual machine introspection (VMI) framework, to get Windows kernel symbols. We chose Volatility since it is a widespread and well-known framework that offers several profiles to inspect the memory of several OSes. Using a popular VMI framework facilitates the portability of Laccolith to new versions of the guest OS (e.g., builds of Microsoft Windows) and new guest OSes since the community continuously provides multiple profiles to align VMI with new releases of the OSes.

\section{Experimental Analysis}
\label{sec:experiment}
We compare Laccolith with MITRE CALDERA since it is a state-of-the-art solution for adversary emulation. Compared to other adversary emulation tools, CALDERA provides greater coverage of APT tactics and techniques, a complete client-server architecture, and the ability to orchestrate complex APT campaigns (see also  Section~\ref{sec:related} for a detailed comparison). We evaluate adversary emulation concerning its \textit{detectability}.

To evaluate detectability, we test the adversary emulation tools against a set of AV solutions, described in Section~\ref{sec:Experimental Setup}. 
As a metric to measure detectability, we consider the \textbf{adversary profile execution progress}. An adversary profile is a sequence of atomic steps that represent malicious activities. The metric represents the progress of the execution of an adversary profile in terms of atomic steps successfully executed until the AV raises an alarm. The \textit{successful execution} of an atomic step means that the agent (from either CALDERA or Laccolith) can execute the action without being detected by the AV. The steps that compose an adversary profile are called \textit{abilities} for CALDERA and \textit{actions} for Laccolith.
This metric is expressed as a fraction:
\begin{displaymath}
    \textrm{Adversary Profile Execution Progress \hspace{3pt}} = \textrm{\hspace{5pt}} \frac{N_{EA}}{N_{PA}} 
\end{displaymath}
$N_{EA}$ is the number of executed actions of an adversary profile, while $N_{PA}$ is the total number of its actions. For example, suppose a profile has 10 actions, and the AV detects the execution of its eighth action. In that case, the adversary profile execution progress will be $7/10$ since the profile managed to execute 7 actions out of 10.

To gain additional insights into the detectability, we also analyze the initial loading of the emulation agent in the victim machine. Since AV products perform rigorous checks before an executable launches, the emulation agent also needs to evade these checks. For this purpose, we introduce an additional metric, the \textbf{injection success}. This metric represents the probability of successfully injecting the agent into the victim without triggering detection. 
The metric is defined as a fraction:
\begin{displaymath}
    \textrm{Injection Success \hspace{3pt}} = \textrm{\hspace{5pt}} \frac{N_{SI}}{N_{I}} 
\end{displaymath}
where $N_{SI}$ is the number of successful injections attempts and $N_{I}$ is the total number of attempts.


Finally, we also include \textit{atomic tools} in our experimental analysis of state-of-the-art, i.e., tools that enable the execution of single (atomic) adversarial actions, outside the context of a complete attack campaign. In particular, we focused on \textit{Atomic Red Team} \cite{AtomicRedTeam} and \textit{Invoke-Adversary} \cite{InvokeAdversary}, two popular atomic toolkits \cite{zilberman2020sok} as discussed in Section \ref{sec:related}. These tools are designed to perform single adversarial actions from the MITRE ATT\&CK matrix. Their atomic nature allows us to overcome the problem of not testing techniques skipped when running an entire campaign with multiple actions, which could still be detected and stopped by the AVs. Specifically, we analyze the percentage of actions that can be detected by AVs.

\subsection{Experimental Setup}
\label{sec:Experimental Setup}

\begin{figure}[!htbp]
\centering
\centerline{\includegraphics[width=1\columnwidth]{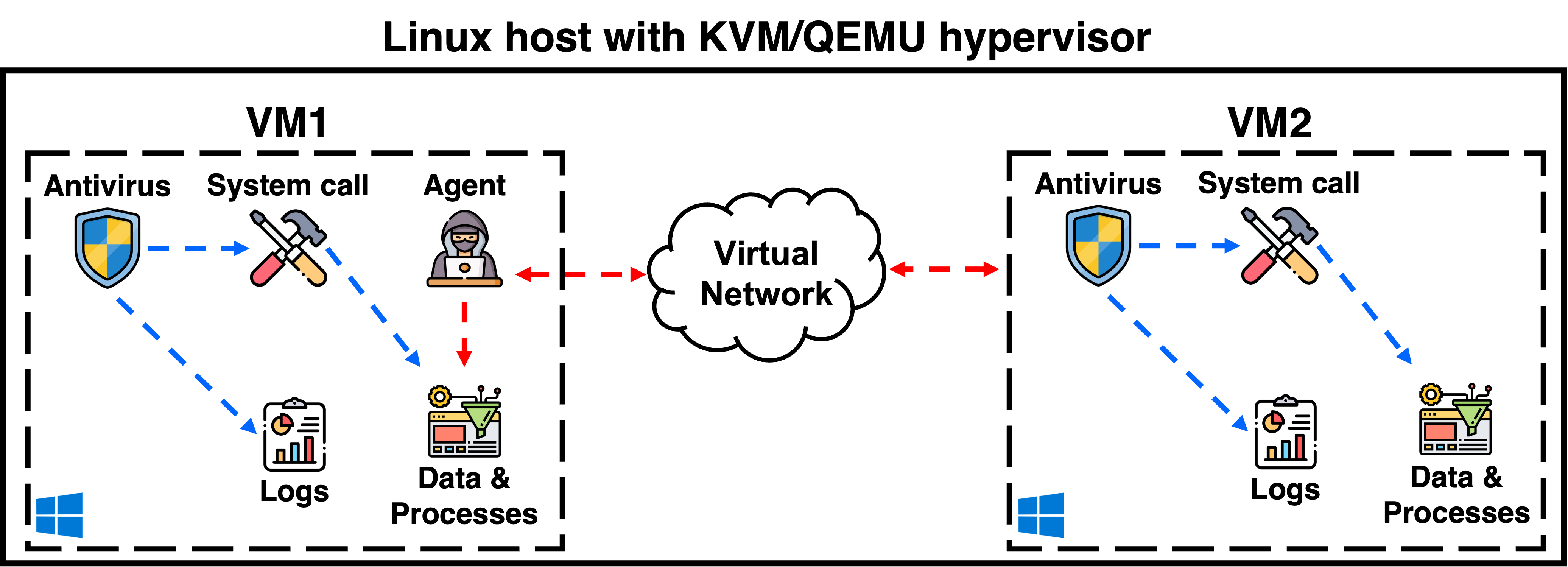}}
\caption{Experimental setup.}
\label{fig:Setup}
\end{figure}

We used the following configuration for the experimental analysis, shown in Figure~\ref{fig:Setup}:
\begin{itemize}
    \item Victim Windows 10 VM (VM1): the target of the adversary emulation;
    \item Additional Windows 10 VM (VM2): it is a machine necessary to perform some malicious actions, e.g., network shares enumeration performed by the \textit{Shares Hunter} profile, as shown in Table \ref{tab:Laccolith profiles}.
\end{itemize}

Both VMs are equipped with 4 CPU cores and 2 GB RAM and run Microsoft Windows 10. Before our experiments, we tested the portability of Laccolith across different versions of the guest OS. We deployed Laccolith on seven Windows 10 versions, which span over three years, with significant changes across the versions \cite{26}. 
Laccolith leverages Volatility to apply the injection method in a portable way. Indeed, Laccolith could correctly inject the emulation agent in all the tested versions of Windows without the need for special customizations. In our experiments, we focus on Microsoft Windows 10 build 19044.



We selected five popular AV products for our detectability analysis: Windows Defender, Avast, AVG, Kaspersky, and Avira. These products qualify as modern EDR solutions and cover state-of-the-art detection techniques \cite{botacin2022antiviruses}. For instance, they monitor system calls using user-space and kernel-space hooks; they use both signature-based and real-time behavioral detection; and they implement self- and system-protection techniques. An example of such protections is preventing filesystem access to specific directories, such as the main system directories and the AV directory itself. 

\subsection{Detectability evaluation of MITRE CALDERA}
To study the detectability of CALDERA, we selected its twelve default adversary profiles. These profiles can be classified into two categories: \textit{reconnaissance \& information gathering} and \textit{advanced}. The first category encompasses the first eight profiles listed below, which perform basic operations: user identification, process enumeration, anti-virus discovery, screenshot capture, and file search. The advanced profiles perform more invasive actions, like process injection, lateral movement, and malicious payload execution. As a consequence, their activities will be noisy and more likely to be flagged by the anti-viruses. We report a short description for each profile:
\begin{itemize}
    \item \textit{Discovery}: collects detailed information from a host, such as local users, user processes, admin shares, and anti-virus programs;
    \item \textit{Hunter}: performs Discovery operations, then tries to exfiltrate files from the working directory;
    \item \textit{Check}: collects information about the configuration of the host (e.g., installation of common software packages, such as Chrome, Go, and Python), and the configuration of its network interfaces;
    \item \textit{Collection}: collects personal information from the host, such as company emails, IP addresses, and personal files;
    \item \textit{Enumerator}: enumerates the presence of different types of processes on the host, such as WMIC, PowerShell, and SysInternals utilities;
    \item \textit{Nosy Neighbor}: finds the preferred Wi-Fi networks, and tries to disrupt the Wi-Fi connection;
    \item \textit{Signed Binary Proxy Execution}: executes malicious code through signed and trusted binaries;
    \item \textit{Super Spy}: monitors the active user by capturing screenshots, copying data from the clipboard, and scanning preferred Wi-Fi networks;
    \item \textit{Undercover}: swaps the built-in PowerShell with PowerShell Core to stop PowerShell processes;
    \item \textit{Stowaway}: injects Sandcat (the default emulation agent in CALDERA) into another process;
    \item \textit{Worm}: runs PowerKatz to steal user credentials, then moves laterally in multiple ways;
    \item \textit{You Shall (Not) Bypass}: bypasses User Account Control (UAC).
\end{itemize}

In addition to these profiles, we implemented a custom \textit{Ransomware} profile for completeness. This profile was developed from scratch, similar to the implementation of the predefined profiles. It encompasses the following abilities:
\begin{itemize}
    \item Find files;
    \item Stage sensitive files;
    \item Compress the staged directory;
    \item Exfiltrate the staged directory;
    \item Encrypt sensitive files (e.g., \textit{docx} and \textit{pdf} files).
\end{itemize}

The first four abilities re-use the ones of the predefined profiles to find the target files. The last ability has been implemented with a PowerShell script \cite{56} to encrypt the sensitive files, by silently skipping the files if it does not have permission to rewrite them. Using a PowerShell script with standard APIs for encryption makes the process more similar to ``legitimate'' programs (as in fileless malware), similar to other profiles in CALDERA. In this way, the Ransomware profile should be as detectable as the other reconnaissance \& information gathering adversary profiles.


\begin{table}[!ht]
\centering
\caption{Adversary Profile Execution Progress for MITRE CALDERA. \textcolor{red}{\textbf{Bold}} values represent incomplete progress.}
\begin{center}
\resizebox{\columnwidth}{!}{%
\begin{tabular}{cccccc}
\hline
\textbf{Profile} & \textbf{Windows Defender} & \textbf{Avast} & \textbf{AVG} & \textbf{Kaspersky} & \textbf{Avira}\\
\hline
Discovery & \DrawPercentageBar{1} \hspace{7pt} 9/9 & \DrawPercentageBar{1} \hspace{7pt} 9/9 & \DrawPercentageBar{1} \hspace{7pt} 9/9 & \DrawPercentageBar{1} \hspace{7pt} 9/9 & \DrawPercentageBar{1} \hspace{7pt} 9/9 \\
\hline
Hunter & \DrawPercentageBar{1} 14/14 & \DrawPercentageBar{1} 14/14 & \DrawPercentageBar{1} 14/14 & \DrawPercentageBar{1} 14/14 & \DrawPercentageBar{1} 14/14 \\
\hline
Check & \DrawPercentageBar{1} \hspace{7pt} 6/6 & \DrawPercentageBar{1} \hspace{7pt} 6/6 & \DrawPercentageBar{1} \hspace{7pt} 6/6 & \DrawPercentageBar{1} \hspace{7pt} 6/6 & \DrawPercentageBar{1} \hspace{7pt} 6/6 \\
\hline
Collection & \DrawPercentageBar{1} \hspace{7pt} 2/2 & \DrawPercentageBar{1} \hspace{7pt} 2/2 & \DrawPercentageBar{1} \hspace{7pt} 2/2 & \DrawPercentageBar{1} \hspace{7pt} 2/2 & \DrawPercentageBar{1} \hspace{7pt} 2/2 \\
\hline
Enumerator & \DrawPercentageBar{1} \hspace{7pt} 5/5 & \DrawPercentageBar{1} \hspace{7pt} 5/5 & \DrawPercentageBar{1} \hspace{7pt} 5/5 & \DrawPercentageBar{1} \hspace{7pt} 5/5 & \DrawPercentageBar{1} \hspace{7pt} 5/5 \\
\hline
Nosy Neighbor & \DrawPercentageBar{1} \hspace{7pt} 7/7 & \DrawPercentageBar{1} \hspace{7pt} 7/7 & \DrawPercentageBar{1} \hspace{7pt} 7/7 & \DrawPercentageBar{1} \hspace{7pt} 7/7 & \DrawPercentageBar{1} \hspace{7pt} 7/7 \\
\hline
Signed Binary Proxy Execution & \DrawPercentageBar{1} \hspace{7pt} 3/3 & \DrawPercentageBar{1} \hspace{7pt} 3/3 & \DrawPercentageBar{1} \hspace{7pt} 3/3 & \DrawPercentageBar{1} \hspace{7pt} 3/3 & \DrawPercentageBar{1} \hspace{7pt} 3/3 \\
\hline
Super Spy & \DrawPercentageBar{1} 11/11 & \DrawPercentageBar{1} 11/11 & \DrawPercentageBar{1} 11/11 & \DrawPercentageBar{1} 11/11 & \DrawPercentageBar{1} 11/11 \\
\hline
Undercover & \DrawPercentageBar{0.5} \hspace{10pt} \textcolor{red}{\textbf{1/2}} & \DrawPercentageBar{0.5} \hspace{10pt} \textcolor{red}{\textbf{1/2}} & \DrawPercentageBar{0.5} \hspace{10pt} \textcolor{red}{\textbf{1/2}} & \DrawPercentageBar{0.5} \hspace{10pt} \textcolor{red}{\textbf{1/2}} & \DrawPercentageBar{1} \hspace{7pt} 2/2 \\
\hline
Stowaway & \DrawPercentageBar{0.5} \hspace{10pt} \textcolor{red}{\textbf{1/2}} & \DrawPercentageBar{0.5} \hspace{10pt} \textcolor{red}{\textbf{1/2}} & \DrawPercentageBar{0.5} \hspace{10pt} \textcolor{red}{\textbf{1/2}} & \DrawPercentageBar{0.5} \hspace{10pt} \textcolor{red}{\textbf{1/2}} & \DrawPercentageBar{1} \hspace{7pt} 2/2 \\
\hline
Worm & \DrawPercentageBar{0.1111} \hspace{10pt} \textcolor{red}{\textbf{1/9}} & \DrawPercentageBar{0.1111} \hspace{10pt} \textcolor{red}{\textbf{1/9}} & \DrawPercentageBar{0.1111} \hspace{10pt} \textcolor{red}{\textbf{1/9}} & \DrawPercentageBar{0.1111} \hspace{10pt} \textcolor{red}{\textbf{1/9}} & \DrawPercentageBar{1} \hspace{7pt} 9/9 \\
\hline
You Shall (Not) Bypass & \DrawPercentageBar{0.5} \hspace{10pt} \textcolor{red}{\textbf{2/4}} & \DrawPercentageBar{0.5} \hspace{10pt} \textcolor{red}{\textbf{2/4}} & \DrawPercentageBar{0.5} \hspace{10pt} \textcolor{red}{\textbf{2/4}} & \DrawPercentageBar{0.25} \hspace{10pt} \textcolor{red}{\textbf{1/4}} & \DrawPercentageBar{0.25} \hspace{10pt} \textcolor{red}{\textbf{1/4}} \\
\hline
Ransomware & \DrawPercentageBar{1} \hspace{7pt} 5/5 & \DrawPercentageBar{1} \hspace{7pt} 5/5 & \DrawPercentageBar{1} \hspace{7pt} 5/5 & \DrawPercentageBar{1} \hspace{7pt} 5/5 & \DrawPercentageBar{1} \hspace{7pt} 5/5 \\
\hline
\end{tabular}%
}
\label{tab:CALDERA profile progress}
\end{center}
\end{table}

Table \ref{tab:CALDERA profile progress} shows the execution progress for each adversary profile of MITRE CALDERA against the five anti-viruses. 
It is possible to notice that the reconnaissance \& information gathering profiles execute without being detected since they do not perform harmful actions. For the advanced profiles, the results of the experiments are quite different. During the execution, all the anti-viruses, except for Avira, flagged the activities of the profiles as suspicious. Moreover, the detected abilities are always the first ones in the atomic order: once an ability has been detected, it is not possible to complete the operation, since the prerequisites for the following ones will not be satisfied. As a consequence, the last abilities of the advanced profiles could not be executed. For instance, the Stowaway profile uses two abilities: since the execution of the first one is stopped by the anti-virus, the second one cannot be tested. Therefore, this profile scored $1/2$ in terms of detectability. We reported the detected abilities for each profile in Table \ref{tab:Detected abilities}. 
This experimental analysis shows that the emulation agent of MITRE CALDERA can stealthily execute most Discovery, Collection, and Exfiltration techniques, but is easily blocked when it tries to perform more intrusive actions, such as Credential Access and Privilege Escalation.

\begin{table*}[ht]
\caption{Abilities of MITRE CALDERA detected by AV products.}
\begin{center}
\begin{tabular}{ccccccc}
\hline
\textbf{Profile} & \textbf{Ability} & \textbf{Windows Defender} & \textbf{Avast} & \textbf{AVG} & \textbf{Kaspersky} & \textbf{Avira}\\
\hline
Undercover & Install PowerShell Core 6 & \ding{51} & \ding{51} & \ding{51} & \ding{51} &\\
\hline
Stowaway & Inject Sandcat into Process & \ding{51} & \ding{51} & \ding{51} & \ding{51} &\\
\hline
Worm & Run PowerKatz & \ding{51} & \ding{51} & \ding{51} & \ding{51} &\\
\hline
You Shall (Not) Bypass & Wow64log DLL Hijack & \ding{51} & \ding{51} & \ding{51} & \ding{51} & \ding{51}\\
\hline
You Shall (Not) Bypass & Bypass UAC Medium & \ding{51} & \ding{51} & \ding{51} & &\\
\hline
\end{tabular}
\label{tab:Detected abilities}
\end{center}
\end{table*}

It is worth noting that the AVs always detect the injection of CALDERA's agent. After performing 20 repetitions of the injection, we conclude that the injection success of CALDERA is $0/20$ against each AV. Consequently, the results in Table \ref{tab:CALDERA profile progress} only refer to the execution of the adversary profiles, assuming that the agent has already been loaded on the target machine.

\subsubsection{Integrating CALDERA with anti-detection}

To overcome the limitations of MITRE CALDERA in terms of anti-detection, we combined it with Inceptor \cite{Inceptor}, a state-of-the-art evasion framework introduced in Section~\ref{sec:background}. 
In this experimental analysis, we focus on Windows Defender, and on the injection stage of the emulation agent of CALDERA (i.e., the initial installation of the agent). We enhance the agent with multiple anti-detection techniques from Inceptor. 

The injection needs to bypass three layers of detection: User Account Control (UAC), signature-based static analysis, and dynamic sandbox analysis. UAC is a protection that involves user interaction, by displaying a message through a GUI frontend (Microsoft SmartScreen \cite{SmartScreen}). Sandbox analysis is a technique used by anti-virus software to analyze potentially malicious files and programs, by running them in a safe and isolated environment. It is particularly useful for detecting new and unknown malware, which can often evade static signature-based detection. In Microsoft Windows, \textit{Windows Defender Application Guard} \cite{ApplicationGuard} is responsible for performing the sandbox analysis, using a virtual container with a specialized version of the Windows OS.

We used the following combination of anti-detection techniques to make the injection stealthier: a native binary template, an encoding chain composed of \textit{Shikata-Ga-Nai} \cite{Shikata}, a popular polymorphic binary encoder \cite{ShikataMandiant}, XOR encoding, 120 seconds of execution delay, and an unhooking technique for EDR bypass. Moreover, we signed the resulting binary with a Microsoft signature using CarbonCopy \cite{41}.

We experimentally evaluate the injection of the CALDERA agent with anti-detection by performing 20 repetitions. The injection is unable to evade UAC. This would require the exploitation of known vulnerabilities in Microsoft Windows (\emph{UAC bypass}), which are regularly fixed by updates of the OS. Therefore, Inceptor is unable to provide a reliable solution to escape UAC, and we manually allow the execution of the CALDERA agent through the UAC. 
Even neglecting the problem with UAC, the agent was successfully injected approximately 5 times out of 20, thus with an injection success equal to $5/20$ and 25\% probability. 
The encoding chain allows the binary to bypass static analysis since it does not match any known signature. The execution delay and unhooking technique heuristically help to elude the dynamic sandbox analysis: the delay is helpful to make the behavior seem benign while unhooking prevents EDRs from inspecting function calls.
The binary signature helped to evade the dynamic analysis: when the binary has a valid signature, Windows Defender performs less detailed dynamic checks. However, since the binary needs to decode itself from the encryption chain at some point, it is still challenging to bypass the sandbox analysis, making the injection detectable. 
The need for additional UAC bypass techniques further complicates the process of making CALDERA stealthier.
\vspace{-10pt}

\subsection{Detectability evaluation of atomic tools}
After CALDERA, we evaluate the detectability of individual actions from atomic tools (\textit{Atomic Red Team} \cite{AtomicRedTeam} and \textit{Invoke-Adversary} \cite{InvokeAdversary}). For the sake of clarity, we focused on testing a subset of the actions provided by these tools. We sampled the actions uniformly across the tactics of the MITRE cyber kill chain. Table~\ref{tab:Detected atomic abilities} provides the percentage of actions detected by the AVs. 
Table \ref{tab:Atomic Red Team results} and \ref{tab:Invoke-Adversary results} provide detailed results for Atomic Red Team and Invoke-Adversary, respectively. These tools show the same drawbacks as CALDERA, where the most intrusive actions all raise alerts from the AVs. Looking at the tables, it is possible to notice how non-intrusive actions are the only ones not detected by any of the AVs, namely \textit{GetCurrent User with PowerShell Script}, \textit{Prompt User for Password}, \textit{Activate Guest Account} for Atomic Red Team, and \textit{System Owner Discovery} and \textit{Screen Capture} for Invoke-Adversary. 
It is worth noting that even the deployment of the tools triggers an alert from the AVs. So, it is not possible to use them without turning AVs off in the deployment phase.

\begin{table}[ht]
\caption{Atomic techniques detection results for each AV product.}
\begin{center}
\resizebox{\columnwidth}{!}{%
\begin{tabular}{cccccc}
\hline
\textbf{Profile} & \textbf{Windows Defender} & \textbf{Avast} & \textbf{AVG} & \textbf{Kaspersky} & \textbf{Avira}\\
\hline
Atomic Red Team & 70\% & 70\% & 70\% & 50\% & 30\%\\
\hline
Invoke-Adversary & 75\% & 75\% & 75\% & 63\% & 50\%\\
\hline
\end{tabular}%
}
\label{tab:Detected atomic abilities}
\end{center}
\end{table}

\begin{table}[!ht]
\centering
\caption{Technique detectability for Atomic Red Team.}
\begin{center}
\resizebox{\columnwidth}{!}{%
\begin{tabular}{cccccc}
\hline
\textbf{Technique} & \textbf{Windows Defender} & \textbf{Avast} & \textbf{AVG} & \textbf{Kaspersky} & \textbf{Avira}\\
\hline
Execute base64-encoded PowerShell from Windows Registry & \ding{52} & \ding{52} & \ding{52} & \ding{52} & \\
\hline
GetCurrent User with PowerShell Script &  &  &  &  &  \\
\hline
Thread Execution Hijacking & \ding{52} & \ding{52} & \ding{52} & \ding{52} & \ding{52} \\                       
\hline
Prompt User for Password &  &  &  &  &  \\
\hline
Mimikatz & \ding{52} & \ding{52} & \ding{52} & \ding{52} & \ding{52} \\
\hline
Clear Logs & \ding{52} & \ding{52} & \ding{52} &  &  \\
\hline
Activate Guest Account &  &  &  &  &  \\
\hline
Disable Windows Security Center Notifications & \ding{52} & \ding{52} & \ding{52} & \ding{52} & \ding{52} \\
\hline
Bypass UAC using Event Viewer (PowerShell) & \ding{52} & \ding{52} & \ding{52} &   & \\
\hline
Disable Microsoft Defender Firewall & \ding{52} & \ding{52} & \ding{52} & \ding{52} &  \\
\hline
\end{tabular}%
}
\label{tab:Atomic Red Team results}
\end{center}
\end{table}

\begin{table}[!ht]
\centering
\caption{Technique detectability for Invoke-Adversary.}
\begin{center}
\resizebox{\columnwidth}{!}{%
\begin{tabular}{cccccc}
\hline
\textbf{Technique} & \textbf{Windows Defender} & \textbf{Avast} & \textbf{AVG} & \textbf{Kaspersky} & \textbf{Avira}\\
\hline
System Owner Discovery &  &  &  &  &  \\
\hline
PowerShell Encoded Mimikatz & \ding{52} & \ding{52} & \ding{52} & \ding{52} &  \ding{52}\\
\hline
Screen Capture &  &  &  &  &  \\
\hline
Add local firewall rule exceptions & \ding{52} & \ding{52} & \ding{52} & \ding{52} &  \\
\hline
Create local administrator & \ding{52} & \ding{52} & \ding{52} & \ding{52} & \ding{52} \\
\hline
Capture Lsass Memory Dump & \ding{52} & \ding{52} & \ding{52} & \ding{52} & \ding{52} \\
\hline
Clear Security Log & \ding{52} & \ding{52} & \ding{52} &  &  \\
\hline
PSExec & \ding{52} & \ding{52} & \ding{52} & \ding{52} & \ding{52} \\
\hline
\end{tabular}%
}
\label{tab:Invoke-Adversary results}
\end{center}
\end{table}
\subsection{Detectability evaluation of Laccolith}

Our previous analysis showed that traditional adversary emulation is cumbersome and unable to perform intrusive actions without being detected. We here present an experimental analysis of detectability for Laccolith. For this analysis, we implemented four adversary profiles with Laccolith: \textit{Thief}, \textit{Op-2}, \textit{Ransomware}, and \textit{Shares Hunter}, as described in Table \ref{tab:Laccolith profiles}. Since we had to develop adversary profiles from scratch for Laccolith, we did not aim for a verbatim reimplementation of the adversary profiles of CALDERA. Instead, the adversary profiles in Laccolith match the CALDERA profiles in terms of high-level strategy of the attackers. Moreover, we relate the adversary profiles for Laccolith with real-world APTs (threat-informed adversary emulation). The new profiles in Laccolith cover all the tactics covered by CALDERA profiles, except Privilege Escalation (covered by the \textit{You Shall (Not) Bypass} profile) since Laccolith already has high privilege. The Ransomware profile also covers the Impact tactic, which is not covered by the default profiles of CALDERA. We introduced this profile to show the ability of Laccolith to support complex operations and cover a relevant cybersecurity threat.

\begin{table*}[ht]
\centering
\caption{Adversary profiles in Laccolith.}
\begin{center}
\resizebox{\textwidth}{!}{%
\begin{tabular}{cccccc}
\hline
\textbf{Profile} & \textbf{Description} & \textbf{Tactics} & \textbf{Commands} & \textbf{High-level actions} & \textbf{Referenced APTs}\\
\hline
& Exfiltrate files & Discovery, & Directory listing (to find & (1) Find local users, & APT1,\\
Thief & from local & Collection, & local users and to list & (2) List user desktop, & OilRig,\\
& user desktop & Exfiltration & desktop files), Read file & (3) Exfiltrate a list of staged files & APT3\\
\hline
& Upload a Powershell script & & & (1) Write file on &\\
& in a system folder & & Write file, & remote file system, &\\
& and install a scheduled & Persistence, & Write to registry, & (2) Install a scheduled & Remsec\\
Op-2 & task that executes that & Credential & Version, & task on the remote & (Strider),\\
& script at boot, get & access & Dump process & Windows target, & Ke3chang\\
& system version and & & memory & (3) Get system version, &\\
& dump LSASS memory & & & (4) Dump lsass credentials &\\
\hline
& & &  & (1) Find local users, &\\
& Discover and exfiltrate & Discovery, & Directory listing, & (2) Find sensitive files, & APT3,\\
Ransomware & sensitive files, encrypt & Collection, & Read file, & (3) Exfiltrate a list of staged & Bad Rabbit\\
& them and leave & Exfiltration, & Write file & files and encrypt them, & (multiple APTs)\\
& a message & Impact & & (4) Encrypt remote files, &\\
& & & & (5) Write ransom message &\\
\hline
& Read ARP cache to find & & & (1) Find local IP address, &\\
& neighbors, scan them to see & Discovery, & User-mode commands: & (2) Read ARP cache, & Conti,\\
Shares Hunter & which has Netbios/SMB & Lateral & ipconfig, arp, & (3) Scan hosts for & APT32\\
& sharing enabled, & Movement & nbtstat, net view & SMB/NetBIOS, &\\
& enumerate shares & & & (4) Enumerate network shares &\\
\hline
\end{tabular}}%
\label{tab:Laccolith profiles}
\end{center}
\end{table*}

Table \ref{tab:Test commands coverage} provides details on which commands of the emulation agent are involved in each profile, in order to get more insight about the adversary profiles of this experiment. In total, the profiles cover 7 commands implemented by the emulation agent, including actions to access the filesystem, the system registry, and to execute user-space commands. Moreover, the Laccolith agent provides additional commands for managing adversary emulation campaigns, not shown in the table for the sake of brevity. Overall, the newly implemented profiles allowed us to test the functionalities of Laccolith in depth.

To compare Laccolith to CALDERA, we performed the same experimental analysis for detectability. 
Table~\ref{tab:Laccolith profile progress} shows the adversary profile execution progress for the four profiles against all the chosen AVs. 
It is possible to notice that all profiles achieve complete execution progress, meaning that the AVs do not detect any of their actions.
It is worth noting that the Shares Hunter profile has seven actions according to Table~\ref{tab:Laccolith profile progress}, instead of the four mentioned in Table~\ref{tab:Laccolith profiles}. This happens because this profile does not have an \textit{a priori} planning, but a dynamic one in which some actions depend on the outcome of the previous ones. In particular, since this profile performs network discovery operations, the result of its actions depends on whether there are any neighbors in the network. We assumed that the target VM interacted with the other VM recently, so it has the IP address of VM2 in its ARP cache alongside the gateway one. Both VM1 and VM2 have the NetBIOS sharing option active.


\begin{table}[ht]
\centering
\caption{Coverage of commands of the Laccolith agent, with respect to the adversary profiles.}
\begin{center}
\resizebox{5cm}{!}{%
\begin{tabular}{ccc}
\hline
\textbf{Profile} & \textbf{Commands}\\
\hline
Thief & dir, read\\
\hline
Op-2 & write, setkey, version, dump\\
\hline
Ransomware & dir, read, write\\
\hline
Shares Hunter & read, usermode\\
\hline
\end{tabular}%
}
\label{tab:Test commands coverage}
\end{center}
\end{table}

\begin{table}[ht]
\centering
\caption{Adversary profile execution progress for Laccolith.}
\begin{center}
\resizebox{\columnwidth}{!}{%
\begin{tabular}{cccccc}
\hline
\textbf{Profile} & \textbf{Windows Defender} & \textbf{Avast} & \textbf{AVG} & \textbf{Kaspersky} & \textbf{Avira}\\
\hline
Thief & \DrawPercentageBar{1} 3/3 & \DrawPercentageBar{1} 3/3 & \DrawPercentageBar{1} 3/3 & \DrawPercentageBar{1} 3/3 & \DrawPercentageBar{1} 3/3 \\
\hline
Op-2 & \DrawPercentageBar{1} 4/4 & \DrawPercentageBar{1} 4/4 & \DrawPercentageBar{1} 4/4 & \DrawPercentageBar{1} 4/4 & \DrawPercentageBar{1} 4/4 \\
\hline
Ransomware & \DrawPercentageBar{1} 5/5 & \DrawPercentageBar{1} 5/5 & \DrawPercentageBar{1} 5/5 & \DrawPercentageBar{1} 5/5 & \DrawPercentageBar{1} 5/5 \\
\hline
Shares Hunter & \DrawPercentageBar{1} 7/7 & \DrawPercentageBar{1} 7/7 & \DrawPercentageBar{1} 7/7 & \DrawPercentageBar{1} 7/7 & \DrawPercentageBar{1} 7/7 \\
\hline
\end{tabular}
}
\label{tab:Laccolith profile progress}
\end{center}
\end{table}




The injection relies on overwriting the code of a system call or a kernel function called by a system call. Then, the shellcode executes only once, and the other concurrent calls return, as described in Section \ref{sec:injection process}. After loading the emulation agent, the original code is restored. This injection method can fail since the original code may be restored concurrently with a thread executing that function, which could encounter invalid opcodes due to misalignment or valid code that returns errors because of invalid values in the CPU registers. This failure results in an assertion failure (e.g., \textit{bug check}) within the kernel. Moreover, a critical system process (e.g., \textit{SYSTEM svchost.exe}) may crash because the system call does not exhibit the expected behavior.


We performed preliminary experiments to gain insights into these events. In these initial experiments, we executed the injection after an increasing amount of time after the boot of the victim machine. We found that these events are more likely if the injection is performed within a few minutes right after the boot. If the injection is performed after a few minutes have passed since boot, the injection becomes more reliable. This behavior can be explained by the higher activity of system processes in the early phases of the start-up, which make high use of system calls and can expose these processes to failures. 

Therefore, the success of the injection method depends on the fraction of time the system spends executing the injected system call, and the percentage of cases that the system call is invoked by system processes. 
To quantify the effectiveness of the injection method, we performed more experiments by focusing on the favorable case of injection after boot has been completed, and the system has stabilized. 
The experiments consisted in repeating the injection method multiple times, each time from a clean condition (e.g., a full restart of the victim machine), and measuring how many times the injection was successful. 
We conducted 20 experiments for each AV, re-booting the target VM each time to have independent samples. The timing of the injection was set to one minute after the Windows login prompt appeared. To verify if the injection was successful, we sent the agent an \textit{echo} and a \textit{close} command to check if it could execute commands and terminate gracefully. 
The injection fails if the connection is received, but it is impossible to perform these operations. If there is some minor error from user applications, which cannot be attributable to a security alarm (e.g., a generic ``unknown exception in explorer.exe''), we still consider the injection successful.

Table~\ref{tab:Empirical reliability} summarizes the experimental results. The AV does not impact the injection success since all the setups with the different AVs exhibit similar behavior. The overall success rate of the injection method has been $90/100$, hence 90\% of experiments. 
Considering a margin of error of $1/\sqrt{N}$, where $N$ is the overall number of repetitions \cite{42}, the success rate of the injection method ranges between 80\% and 100\%. Even in the worst case, the success rate is still higher than the probability of running CALDERA and Inceptor without being detected, which had approximately a 25\% probability of success, neglecting the issue of bypassing UAC which makes the process even more uncertain.

It is worth noting that the injection success for Laccolith can be further improved by injecting into a target linear region of code that is executed less frequently, in order to reduce the probability that the injection clashes with the execution of the target system call. This can be achieved by a preliminary profiling of the execution frequency of system calls. Since the profiling depends on the workload of the specific system under evaluation, and since this represents an engineering problem, we consider this beyond the scope of our research work. 

\begin{table}[ht]
\centering
\caption{Injection success for Laccolith.}
\begin{center}
\resizebox{4cm}{!}{%
\begin{tabular}{cc}
\toprule
\textbf{Anti-virus} & \textbf{Injection Success} \\
\toprule
Windows Defender & \DrawPercentageBar{0.95} \hspace{2pt} 19/20 \\
\hline
Avast & \DrawPercentageBar{0.85} \hspace{2pt} 17/20 \\
\hline
AVG & \DrawPercentageBar{0.85} \hspace{2pt} 17/20 \\
\hline
Kaspersky & \DrawPercentageBar{0.95} \hspace{2pt} 19/20 \\
\hline
Avira & \DrawPercentageBar{0.90} \hspace{2pt} 18/20 \\
\hline
\hline
Overall & \DrawPercentageBar{0.90} 90/100 \\
\bottomrule
\end{tabular}%
}
\label{tab:Empirical reliability}
\end{center}
\end{table}
\subsection{Threats to validity}
\textbf{Threats to external validity.} We targeted Microsoft Windows as \textbf{guest OS} for the victim machine, which may affect the generality of our solution. 
We focus on Windows since it is a common target for APTs and is widely used in the adversary emulation landscape. For example, the MITRE ATT\&CK framework started as a project to gather information about TTPs against Windows-based systems \cite{strom2018mitre}. It exhibits a diverse attack surface, encompassing a wide range of services, applications, and configurations, which enables a comprehensive exploration of attack vectors. This complexity makes Windows a realistic target for enterprise systems. Other guest OSes can also be targeted by our solution, as discussed later in the paper.
We performed experiments with \textbf{three state-of-the-art open-source adversary emulation tools}: MITRE CALDERA, Atomic Red Team, and Invoke-Adversary. 
According to a recent survey \cite{zilberman2020sok}, these tools are the most mature and are aligned to the MITRE ATT\&CK matrix, which allow us to reproduce threat behavior in a reliable way. 
We evaluated the detectability of adversary emulation against five popular AV products. The \textbf{choice of antivirus solutions} can be a limitation of our evaluation. We focused on antivirus solutions that represent those commonly used in real-world scenarios \cite{botacin2022antiviruses}, considering the ones that aligned better with the typical deployment context of adversary emulation. 

\vspace{3pt}
\noindent
\textbf{Threats to internal validity.} \textbf{Inconsistent configuration settings for the AVs} may threaten the internal validity of the study. We used standard configurations of the five AV solutions, to avoid variations of results not due to the effectiveness of the adversary emulation tools.

\vspace{3pt}
\noindent
\textbf{Threats to construct validity.} The \textbf{choice of detection metrics} can impact the validity of the study. We adopted three different metrics for the evaluation of detectability, which encompass all the key aspects of detectability: perimeter breach detection (\textit{injection success}), malicious actions detection (\textit{atomic technique detection}), and timeliness in the APT campaign detection (\textit{adversary profile execution progress}).
\section{Related Work}
\label{sec:related}

\begin{table*}[ht!]
\caption{Adversary emulation solutions comparison.}
\label{tab:Tools comparison}
\resizebox{\textwidth}{!}{%
\footnotesize 
{\color{black}\begin{tabular}[h!]{ccccccc}
\toprule
\normalsize\textbf{Tool}  & 
\normalsize\textbf{C2 Server} &
\normalsize\textbf{Complex Attacks} &
\normalsize\textbf{ATT\&CK Tactics Coverage} &
\normalsize\textbf{Needs Pre-Installed Agent} &

\normalsize\textbf{Anti-detection} \\ \toprule

CALDERA & 
    \normalsize\ding{51} 
    & 
    \includegraphics[width=0.065\columnwidth]{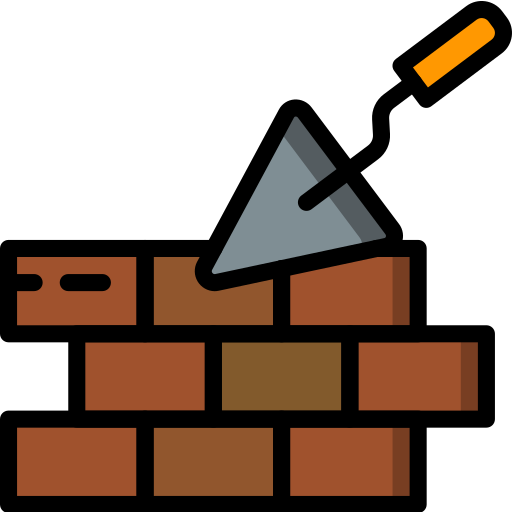}
    \includegraphics[width=0.06\columnwidth]{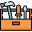} 
    &
    \includegraphics[width=0.065\columnwidth]{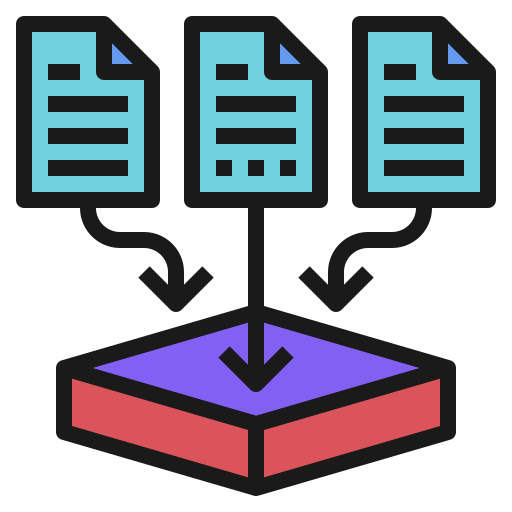} 
    \includegraphics[width=0.065\columnwidth]{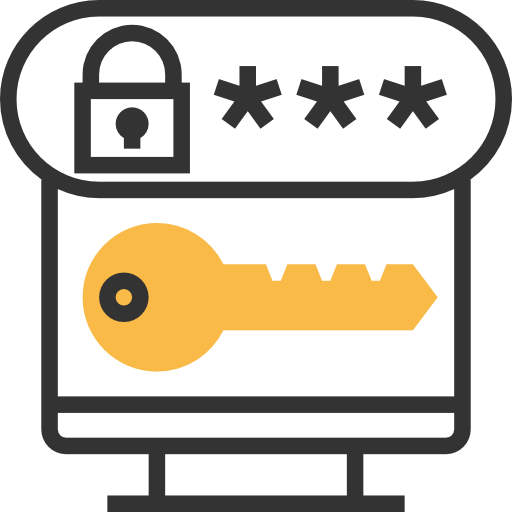}
    \includegraphics[width=0.065\columnwidth]{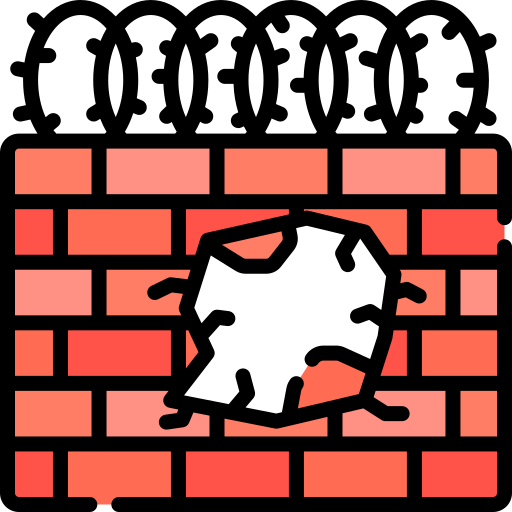}
    \includegraphics[width=0.065\columnwidth]{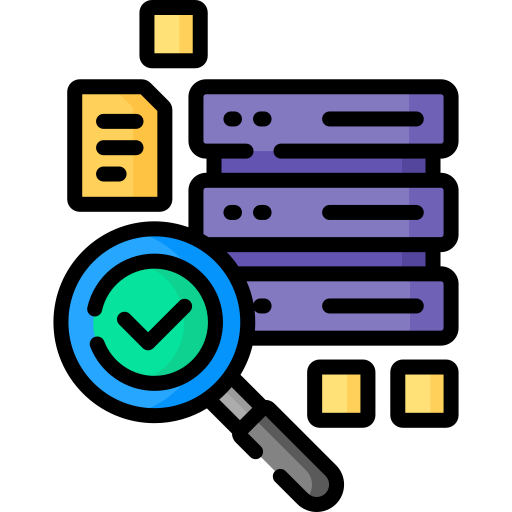} 
    \includegraphics[width=0.065\columnwidth]{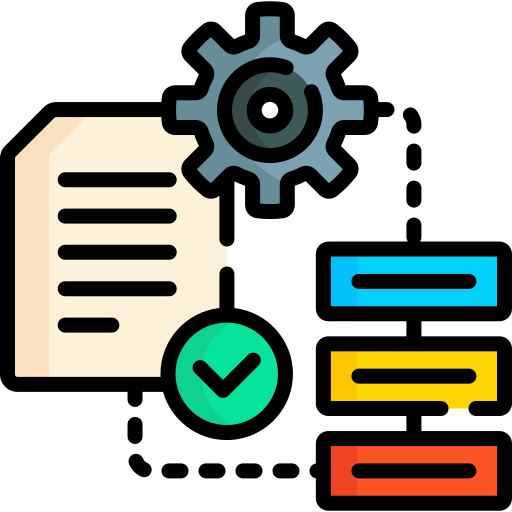} 
    \includegraphics[width=0.065\columnwidth]{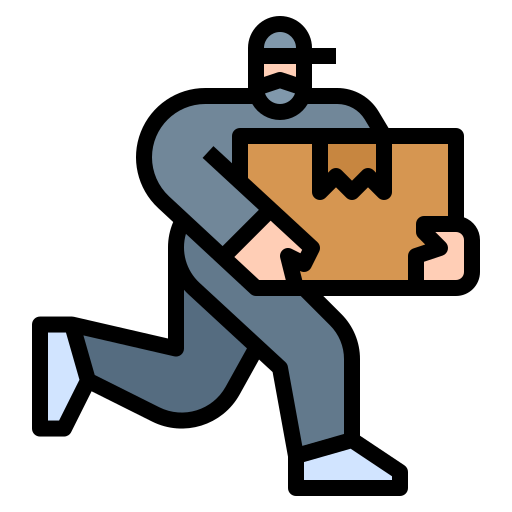} 
    \includegraphics[width=0.065\columnwidth]{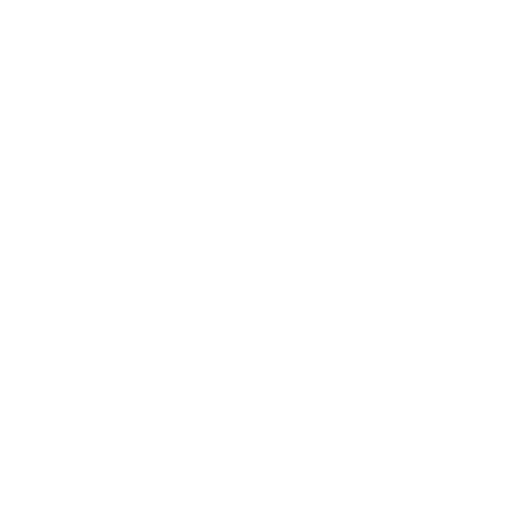}
    \includegraphics[width=0.065\columnwidth]{img/table/tactics/Void.png}
    \includegraphics[width=0.065\columnwidth]{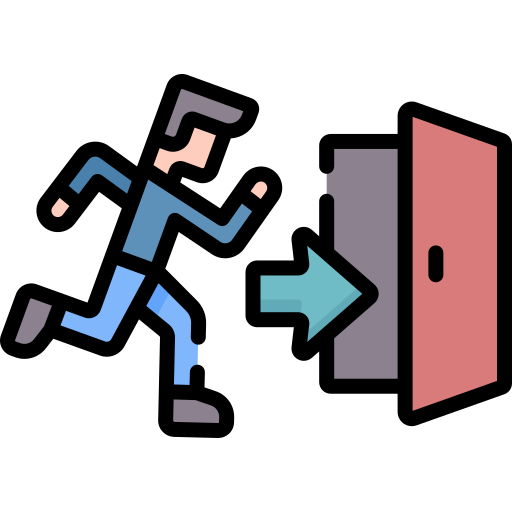}
    \includegraphics[width=0.065\columnwidth]{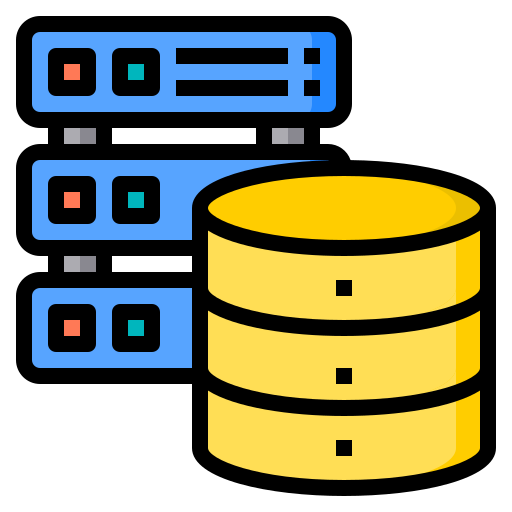} 
    \includegraphics[width=0.065\columnwidth]{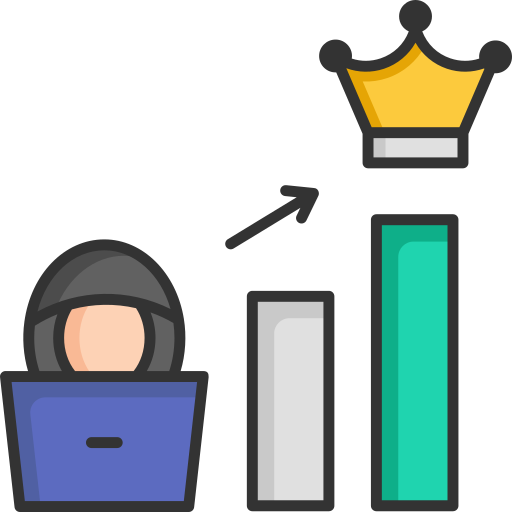}
     
    &
    \normalsize\ding{51}
    &
    \normalsize\ding{55}
    \\ \midrule

Atomic Red Team & 
    \normalsize\ding{55} 
    &  
    \normalsize\ding{55} 
    &
    \includegraphics[width=0.065\columnwidth]{img/table/tactics/Collection.png} 
    \includegraphics[width=0.065\columnwidth]{img/table/tactics/CredentialAccess.png}
    \includegraphics[width=0.065\columnwidth]{img/table/tactics/DefenseEvasion.png}
    \includegraphics[width=0.065\columnwidth]{img/table/tactics/Discovery.png} 
    \includegraphics[width=0.065\columnwidth]{img/table/tactics/Execution.png} 
    \includegraphics[width=0.065\columnwidth]{img/table/tactics/Exfiltration.png} 
    \includegraphics[width=0.065\columnwidth]{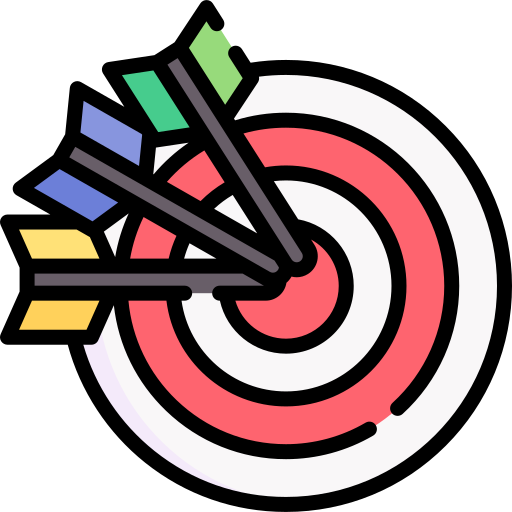} 
    \includegraphics[width=0.065\columnwidth]{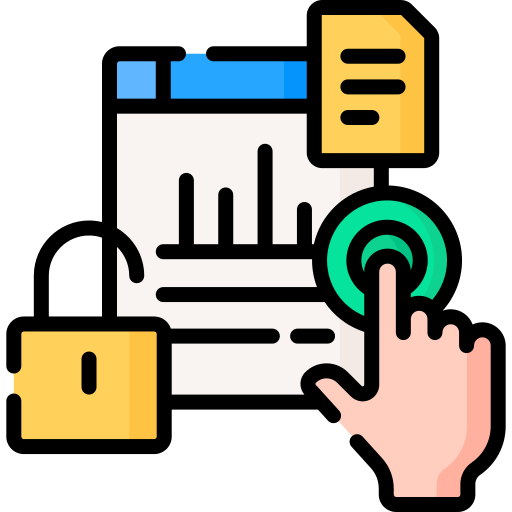}
    \includegraphics[width=0.065\columnwidth]{img/table/tactics/LateralMovement.png}
    \includegraphics[width=0.065\columnwidth]{img/table/tactics/Persistence.png} 
    \includegraphics[width=0.065\columnwidth]{img/table/tactics/PrivilegeEscalation.png} 
    &
    \normalsize\ding{55}
    &
    \normalsize\ding{55}
    \\ \midrule

Red Team Automation & 
    \normalsize\ding{55} 
    & 
    \includegraphics[width=0.065\columnwidth]{img/table/Built_in.png}
    \includegraphics[width=0.06\columnwidth]{img/table/Custom.png} 
    &
    \includegraphics[width=0.065\columnwidth]{img/table/tactics/Void.png} 
    \includegraphics[width=0.065\columnwidth]{img/table/tactics/Void.png}
    \includegraphics[width=0.065\columnwidth]{img/table/tactics/DefenseEvasion.png}
    \includegraphics[width=0.065\columnwidth]{img/table/tactics/Discovery.png} 
    \includegraphics[width=0.065\columnwidth]{img/table/tactics/Execution.png} 
    \includegraphics[width=0.065\columnwidth]{img/table/tactics/Void.png}
    \includegraphics[width=0.065\columnwidth]{img/table/tactics/Void.png}
    \includegraphics[width=0.065\columnwidth]{img/table/tactics/Void.png}
    \includegraphics[width=0.065\columnwidth]{img/table/tactics/LateralMovement.png}
    \includegraphics[width=0.065\columnwidth]{img/table/tactics/Persistence.png} 
    \includegraphics[width=0.065\columnwidth]{img/table/tactics/PrivilegeEscalation.png}
     
    &
    \normalsize\ding{55}
    &
    \normalsize\ding{55}
    \\ \midrule

APTSimulator & 
    \normalsize\ding{55} 
    &
    \includegraphics[width=0.065\columnwidth]{img/table/Built_in.png}
    \includegraphics[width=0.06\columnwidth]{img/table/Custom.png} 
    &
    \includegraphics[width=0.065\columnwidth]{img/table/tactics/Collection.png} 
    \includegraphics[width=0.065\columnwidth]{img/table/tactics/CredentialAccess.png}
    \includegraphics[width=0.065\columnwidth]{img/table/tactics/DefenseEvasion.png}
    \includegraphics[width=0.065\columnwidth]{img/table/tactics/Discovery.png} 
    \includegraphics[width=0.065\columnwidth]{img/table/tactics/Execution.png} 
    \includegraphics[width=0.065\columnwidth]{img/table/tactics/Void.png} 
    \includegraphics[width=0.065\columnwidth]{img/table/tactics/Void.png}
    \includegraphics[width=0.065\columnwidth]{img/table/tactics/Void.png}
    \includegraphics[width=0.065\columnwidth]{img/table/tactics/Void.png}
    \includegraphics[width=0.065\columnwidth]{img/table/tactics/Persistence.png}  
    \includegraphics[width=0.065\columnwidth]{img/table/tactics/Void.png}
     
    &
    \normalsize\ding{55}
    &
    \normalsize\ding{55}
    \\ \midrule

Infection Monkey & 
    
    \normalsize\ding{51} 
    &
    \includegraphics[width=0.065\columnwidth]{img/table/Built_in.png}
    \includegraphics[width=0.065\columnwidth]{img/table/tactics/Void.png}
    &
    \includegraphics[width=0.065\columnwidth]{img/table/tactics/Collection.png} 
    \includegraphics[width=0.065\columnwidth]{img/table/tactics/CredentialAccess.png}
    \includegraphics[width=0.065\columnwidth]{img/table/tactics/DefenseEvasion.png}
    \includegraphics[width=0.065\columnwidth]{img/table/tactics/Discovery.png} 
    \includegraphics[width=0.065\columnwidth]{img/table/tactics/Execution.png} 
    \includegraphics[width=0.065\columnwidth]{img/table/tactics/Exfiltration.png} 
    \includegraphics[width=0.065\columnwidth]{img/table/tactics/Void.png}
    \includegraphics[width=0.065\columnwidth]{img/table/tactics/Void.png}
    \includegraphics[width=0.065\columnwidth]{img/table/tactics/LateralMovement.png}
    \includegraphics[width=0.065\columnwidth]{img/table/tactics/Persistence.png} 
    \includegraphics[width=0.065\columnwidth]{img/table/tactics/PrivilegeEscalation.png}
     
    &
    \normalsize\ding{51}
    &
    \normalsize\ding{55}
    \\ \midrule

Metta & 
    \normalsize\ding{55} 
    &
    \includegraphics[width=0.065\columnwidth]{img/table/Built_in.png}
    \includegraphics[width=0.065\columnwidth]{img/table/tactics/Void.png}
    &
    \includegraphics[width=0.065\columnwidth]{img/table/tactics/Collection.png} 
    \includegraphics[width=0.065\columnwidth]{img/table/tactics/CredentialAccess.png}
    \includegraphics[width=0.065\columnwidth]{img/table/tactics/DefenseEvasion.png}
    \includegraphics[width=0.065\columnwidth]{img/table/tactics/Discovery.png} 
    \includegraphics[width=0.065\columnwidth]{img/table/tactics/Execution.png} 
    \includegraphics[width=0.065\columnwidth]{img/table/tactics/Exfiltration.png}
    \includegraphics[width=0.065\columnwidth]{img/table/tactics/Void.png}
    \includegraphics[width=0.065\columnwidth]{img/table/tactics/Void.png}
    \includegraphics[width=0.065\columnwidth]{img/table/tactics/Void.png}
    \includegraphics[width=0.065\columnwidth]{img/table/tactics/Void.png}
    \includegraphics[width=0.065\columnwidth]{img/table/tactics/Void.png}
    
    &
    \normalsize\ding{55}
    &
    \normalsize\ding{55}
    \\ \midrule

DumpsterFire & 
    \normalsize\ding{55} &

    \includegraphics[width=0.065\columnwidth]{img/table/Built_in.png}
    \includegraphics[width=0.06\columnwidth]{img/table/Custom.png} 
    &
    \includegraphics[width=0.065\columnwidth]{img/table/tactics/Collection.png} 
    \includegraphics[width=0.065\columnwidth]{img/table/tactics/CredentialAccess.png}
    \includegraphics[width=0.065\columnwidth]{img/table/tactics/Void.png}
    \includegraphics[width=0.065\columnwidth]{img/table/tactics/Discovery.png} 
    \includegraphics[width=0.065\columnwidth]{img/table/tactics/Void.png}
    \includegraphics[width=0.065\columnwidth]{img/table/tactics/Exfiltration.png} 
    \includegraphics[width=0.065\columnwidth]{img/table/tactics/Void.png}
    \includegraphics[width=0.065\columnwidth]{img/table/tactics/Void.png}
    \includegraphics[width=0.065\columnwidth]{img/table/tactics/Void.png}
    \includegraphics[width=0.065\columnwidth]{img/table/tactics/Persistence.png} 
    \includegraphics[width=0.065\columnwidth]{img/table/tactics/PrivilegeEscalation.png} 
    &
    \normalsize\ding{55}
    &
    \normalsize\ding{55}
    \\ \midrule
    
Invoke-Adversary & 
    \normalsize\ding{55} 
    &
    \normalsize\ding{55}
    &
    \includegraphics[width=0.065\columnwidth]{img/table/tactics/Void.png}
    \includegraphics[width=0.065\columnwidth]{img/table/tactics/CredentialAccess.png}
    \includegraphics[width=0.065\columnwidth]{img/table/tactics/DefenseEvasion.png}
    \includegraphics[width=0.065\columnwidth]{img/table/tactics/Discovery.png} 
    \includegraphics[width=0.065\columnwidth]{img/table/tactics/Execution.png} 
    \includegraphics[width=0.065\columnwidth]{img/table/tactics/Void.png} 
    \includegraphics[width=0.065\columnwidth]{img/table/tactics/Void.png} 
    \includegraphics[width=0.065\columnwidth]{img/table/tactics/Void.png}
    \includegraphics[width=0.065\columnwidth]{img/table/tactics/Void.png}
    \includegraphics[width=0.065\columnwidth]{img/table/tactics/Persistence.png} 
    \includegraphics[width=0.065\columnwidth]{img/table/tactics/Void.png}
     
    &
    \normalsize\ding{55}
    &
    \normalsize\ding{55}
    \\ \midrule

Sliver & 
    \normalsize\ding{51} 
    &
    \normalsize\ding{55}
    &
    \includegraphics[width=0.065\columnwidth]{img/table/tactics/Collection.png} 
    \includegraphics[width=0.065\columnwidth]{img/table/tactics/Void.png}
    \includegraphics[width=0.065\columnwidth]{img/table/tactics/DefenseEvasion.png}
    \includegraphics[width=0.065\columnwidth]{img/table/tactics/Discovery.png} 
    \includegraphics[width=0.065\columnwidth]{img/table/tactics/Void.png}
    \includegraphics[width=0.065\columnwidth]{img/table/tactics/Exfiltration.png} 
    \includegraphics[width=0.065\columnwidth]{img/table/tactics/Void.png}
    \includegraphics[width=0.065\columnwidth]{img/table/tactics/Void.png}
    \includegraphics[width=0.065\columnwidth]{img/table/tactics/Void.png}
    \includegraphics[width=0.065\columnwidth]{img/table/tactics/Void.png}
    \includegraphics[width=0.065\columnwidth]{img/table/tactics/PrivilegeEscalation.png}  
     
    &
    \normalsize\ding{51}
    &
    \normalsize\ding{55}
    \\ \midrule\midrule
    
\textbf{Laccolith} & 
    \normalsize\ding{51} 
    &
    \includegraphics[width=0.065\columnwidth]{img/table/Built_in.png}
    \includegraphics[width=0.06\columnwidth]{img/table/Custom.png} 
    &
    \includegraphics[width=0.065\columnwidth]{img/table/tactics/Collection.png} 
    \includegraphics[width=0.065\columnwidth]{img/table/tactics/CredentialAccess.png}
    \includegraphics[width=0.065\columnwidth]{img/table/tactics/DefenseEvasion.png}
    \includegraphics[width=0.065\columnwidth]{img/table/tactics/Discovery.png} 
    \includegraphics[width=0.065\columnwidth]{img/table/tactics/Execution.png} 
    \includegraphics[width=0.065\columnwidth]{img/table/tactics/Exfiltration.png} 
    \includegraphics[width=0.065\columnwidth]{img/table/tactics/Impact.png} 
    \includegraphics[width=0.065\columnwidth]{img/table/tactics/InitialAccess.png}
    \includegraphics[width=0.065\columnwidth]{img/table/tactics/LateralMovement.png}
    \includegraphics[width=0.065\columnwidth]{img/table/tactics/Persistence.png} 
    \includegraphics[width=0.065\columnwidth]{img/table/tactics/PrivilegeEscalation.png} 

    &
    \normalsize\ding{55}
    &
    \normalsize\ding{51}
    \\ \bottomrule
    
\end{tabular}}
}%

\includegraphics[width=1.25\columnwidth]{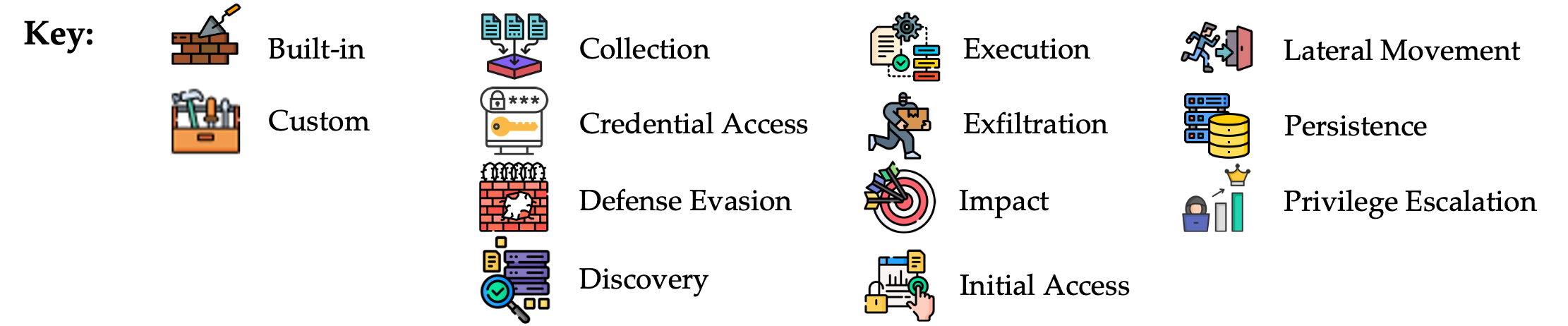}

\end{table*}

Zilberman \emph{et al.} \cite{zilberman2020sok} analyze and classify several adversary emulators. The authors defined a set of criteria and a methodology to evaluate the threat emulators; a taxonomy of the qualities of the threat emulators; and guidelines for choosing an appropriate adversary emulator, given the specific environment and the security assessment tasks. Among the criteria defined in this work are OS compatibility, changes needed in the security array, ATT\&CK TTPs coverage, procedures configuration, and required security expertise.
In this section, we further analyze and discuss the most popular and mature adversary emulation tools according to the survey \cite{zilberman2020sok}.

\emph{Atomic Red Team} \cite{AtomicRedTeam} is a library of scripts to emulate adversary behavior. Every script implements a single ATT\&CK technique or sub-technique (277 out of 719, 231 out of 507 for Windows). It can be used for specific/atomic tests but is not suitable for emulating complex scenarios.
\emph{Red Team Automation} \cite{RedTeamAutomation} is a script framework that implements single techniques from the ATT\&CK framework (around 50) for security assessment purposes. It does not offer built-in scripts for multi-procedure attacks. 
\emph{APTSimulator} \cite{APTSimulator} is a batch script-based tool for Windows that offers around 30 attack techniques to emulate post-compromise scenarios. These techniques leverage external tools such as Mimikatz and PowerSploit. 
\emph{Infection Monkey} \cite{InfectionMonkey} is an adversary emulation tool composed of two main elements, the server (Monkey Island) and the agents (Monkeys). Infection Monkey implements a few ATT\&CK techniques, mostly for Initial Access and Lateral Movement. As CALDERA, Infection Monkey does not offer any anti-detection capability to hide malicious actions, and AVs easily detect its agents' activities \cite{zilberman2020sok}. 
\emph{Metta} \cite{Metta} is an adversary emulator developed by Uber Technologies to assess endpoint security. It provides techniques for various tactics, such as Discovery, Credential Access, and Defense Evasion. However, it does not emulate complete attacks but focuses on the assessment of specific targets. 
\emph{DumpsterFire} \cite{DumpsterFire} offers a collection of actions (fires) that can be chained together into complex attacks (dumpster fires). Fires are not based on ATT\&CK framework techniques and are organized into categories: Network Scans, File Downloads, Websurfing, Account Bruteforcing, Filesystem Activities, Malware, Custom OS Commands, and Shenanigans. It can only execute on Linux.
\emph{Invoke-Adversary} \cite{InvokeAdversary} is a PowerShell script to test security mechanisms. It offers 39 techniques grouped by ATT\&CK tactics. It does not offer multi-procedure scripts and does not support Lateral Movement. Moreover, its procedures are usually detected by AVs, making it not suitable for usage in real-world scenarios.
\emph{Sliver} \cite{Sliver} is a C2 framework for adversary emulation. It offers multi-platform agents that communicate with the C2 server using different protocols (e.g., HTTP, DNS, Mutual TLS). It also provides attacker capabilities through plugins called armories. Sliver does not offer any AV-evasion capability as explicitly stated in its documentation. 

Table~\ref{tab:Tools comparison} summarizes the comparison among the mentioned adversary emulation solutions. It is possible to notice how the tools that provide an advanced architecture with a C2 server (i.e., CALDERA, Infection Monkey, and Sliver) need to install an agent on the target system. None of the other tools, which do not use an agent, offer command-and-control capabilities. In terms of ATT\&CK tactics coverage, Atomic Red Team is the best-performing tool, covering 11 tactics. CALDERA has a similar coverage to Atomic Red Team and is the most complete among the C2 frameworks. CALDERA does not cover Initial Access, because it is not relevant in adversary emulation, and the Impact tactic, unlike our solution which encompasses it in the Ransomware profile. 
Moreover, CALDERA surpasses the other C2 frameworks regarding complex attacks, providing built-in attacks and capabilities to develop custom ones. CALDERA also provides a plugin to leverage Atomic Red Team tactics and techniques, making it the best-performing tool among the existing ones. It is worth noting that none of the analyzed tools provide anti-detection techniques to evade AV/EDR products.

Our proposed solution differs from the previous ones. Laccolith represents a C2 framework that does not need to explicitly install an agent on the victim machine, unlike CALDERA, Infection Monkey, and Sliver. It also offers profiles based on real-world APTs and mapped to the ATT\&CK matrix,  covering the same tactics of Atomic Red Team. Most importantly, Laccolith enables the emulation of malicious actions without being detected by AV and EDR solutions. Moreover, it is portable across different versions of the target guest OS and does not require any customization related to the specific AV/EDR. To achieve all these abilities, Laccolith leverages virtualization, widely used in infrastructures for cybersecurity exercises, as illustrated in the following.
Yamin \emph{et al.} \cite{yamin2020cyber} performed a literature review on Cyber Range platforms and security testbeds, highlighting that most solutions leverage virtualization to set up security training and assessment environments. The use of virtualization spans multiple domains, such as Internet-of-Things (IoT), autonomous systems, SCADA systems, and critical infrastructures.
Beuran \emph{et al.} \cite{beuran2018integrated} proposed \emph{CyTrONE}, a framework for cybersecurity training. CyTrONE is equipped with a Learning Management System (LMS) for the trainees and offers a Cyber Range that relies on virtualization technologies. The Cyber Range is instantiated using a variable number of VMs, depending on the specific training scenario.
\emph{KYPO} \cite{vceleda2015kypo} is a platform for cyber defense exercises developed by Masaryk University. It aims to emulate attacks on critical infrastructures in a controlled environment. For this purpose, KYPO leverages cloud computing technologies (e.g., OpenStack), and relies on virtualization for hosts and networks.
\emph{DETERlab} \cite{wroclawski2016deterlab} is a solution for cybersecurity experimentation developed in the context of the \emph{DETER} project. It provides a flexible and realistic environment for security training and assessment. These goals are fulfilled by leveraging virtualization technologies, which enable easy reconfiguration and scalability. 
The \emph{National Cyber Range} (NCR) \cite{ferguson2014national} is a facility for cybersecurity testing established by the Defense Advanced Research Projects Agency (DARPA). Its purpose is to provide an environment to design and test new ways to respond to the most recent threat actors. The range combines physical and virtualized resources and networks according to the nature of the simulation.

\section{Conclusion}
\label{sec:conclusion}
In this work, we presented Laccolith, a hypervisor-based solution for adversary emulation equipped with anti-detection capabilities. Laccolith introduces a new approach to inject an agent into the target machine based on virtualization, an enabling technology for cybersecurity assessment and training infrastructures.
Laccolith supports the emulation of advanced cyber-attacks (APTs) that adopt sophisticated techniques to hide their traces. This architecture enables the development of a new generation of cyber-range simulations for critical domains, such as in defense. Moreover, Laccolith can support future research on solutions to detect stealthy adversaries.
We performed an experimental analysis to compare Laccolith with state-of-the-art tools for adversary emulation. Our analysis showed that existing tools cannot evade detection by AV/EDR solutions, limiting the usefulness of adversary emulation. We also integrated anti-detection solutions, namely Inceptor, to find out that this combination was still unable to evade defense mechanisms. Instead, Laccolith evaded all the AVs with high reliability, making it suitable for usage in actual emulation scenarios.

A limitation of this work is the focus on a specific guest OS (the victim) and hypervisor. Currently, our prototype supports Microsoft Windows as guest OS and Linux-KVM as hypervisor, since they represent popular solutions for enterprise systems. For the portability of our solution to a different hypervisor, we need APIs from the hypervisor to read and modify the memory of the guest OS. For example, in the case of the XEN hypervisor, the LibVMI library can be leveraged by our solution. If such APIs are not available for the target hypervisor, an alternative is to save and modify a snapshot of the VM memory. For the portability to a different guest OS, we need to rewrite the injector to use kernel-level APIs. This can be easily achieved by taking advantage of the modular architecture of modern OSes, such as, using the APIs for Linux Kernel Modules.
About the scalability, our current prototype requires a few minutes for injecting into VMs with a small amount of RAM. For systems with more RAM, the performance can be improved by optimizing the current prototype. In particular, the most expensive step in the injection process is the analysis of memory through Virtual Machine Introspection (VMI). Our prototype leverages Volatility for VMI, which collects large amounts of information that may be unnecessary for the injection. Therefore, our prototype can be further optimized by tailoring Volatility for the injection phase. 
\section*{Acknowledgments}
This work has been partially supported by the Italian Ministry of University and Research (MUR) under the programme ``PON Ricerca e Innovazione 2014-2020 – Dottorati innovativi con caratterizzazione industriale'' and by MUR PRIN 2022, project FLEGREA, CUP E53D23007950001 (\url{https://flegrea.github.io}).


%

    





\ifCLASSOPTIONcaptionsoff
  \newpage
\fi



%


\bibliographystyle{IEEEtran}
\bibliography{biblio}

%
\begin{IEEEbiography}[{\includegraphics[width=1in,height=1.25in,clip,keepaspectratio]{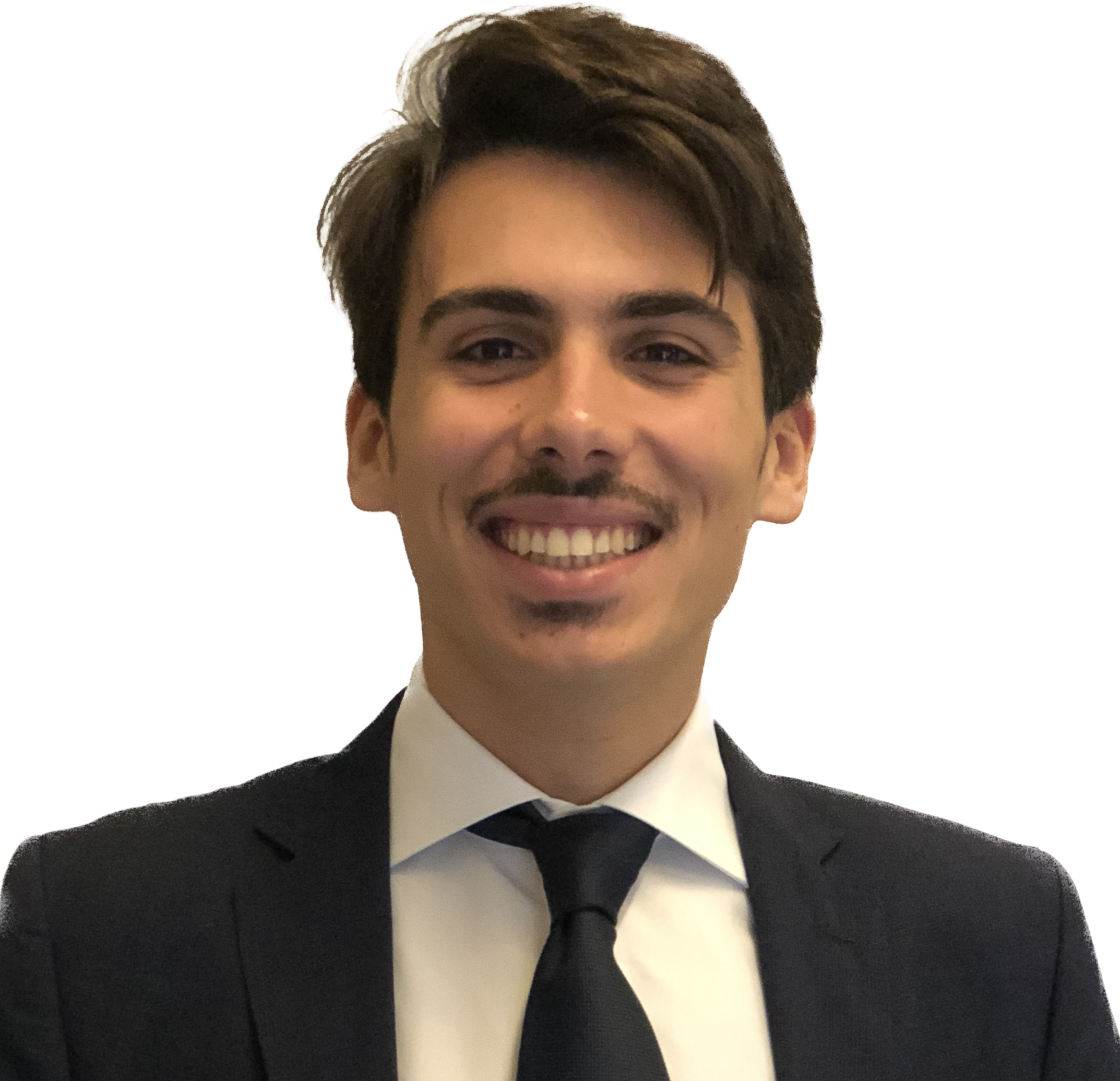}}]{Vittorio Orbinato} (Ph.D.) is a Research Fellow at Università degli Studi di Napoli Federico II, Naples, Italy. His research interests include adversary emulation and software security. He received his Ph.D. from Università degli Studi di Napoli Federico II, Naples, Italy.
\end{IEEEbiography}

\begin{IEEEbiography}[{\includegraphics[width=1in,height=1.25in,clip,keepaspectratio]{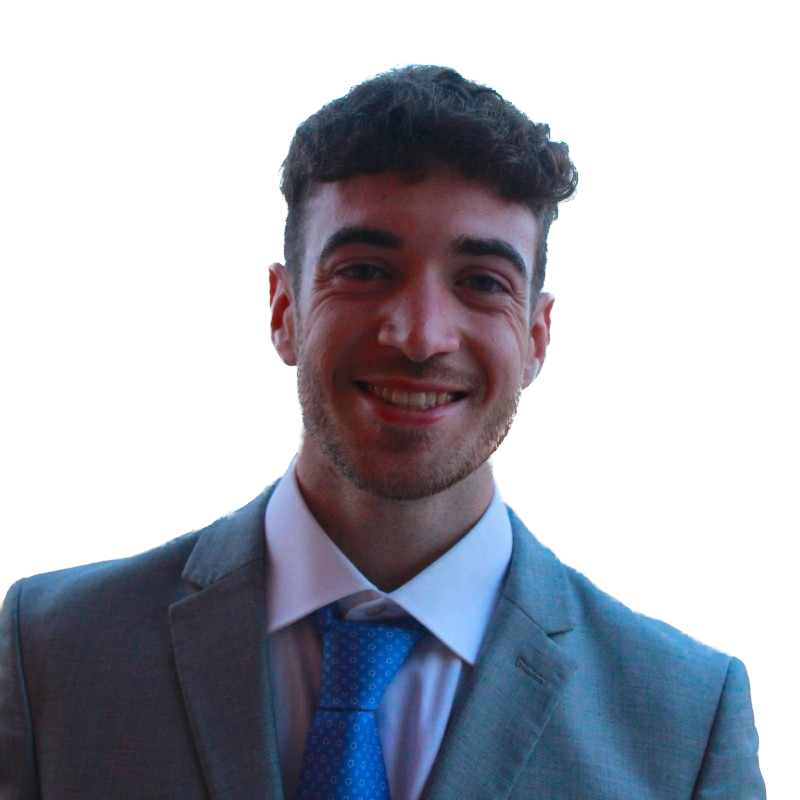}}]{Marco Carlo Feliciano} is a Security Researcher at Secureware s.r.l. His research interests are in adversary emulation. He received his M.Sc. Degree from Università degli Studi di Napoli Federico II, Naples, Italy.
\end{IEEEbiography}

\begin{IEEEbiography}[{\includegraphics[width=1in,height=1.25in,clip,keepaspectratio]{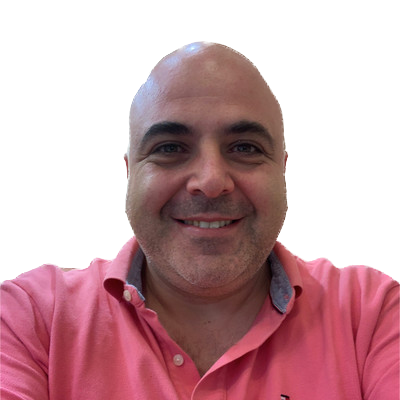}}]{Domenico Cotroneo} (Ph.D.) is Full Professor at Università degli Studi di Napoli Federico II, Naples, Italy. His research interests include software fault injection, dependability assessment, and field-based measurement techniques. He is co-founder and scientific consultant at Secureware s.r.l.
\end{IEEEbiography}

\begin{IEEEbiography}[{\includegraphics[width=1in,height=1.25in,clip,keepaspectratio]{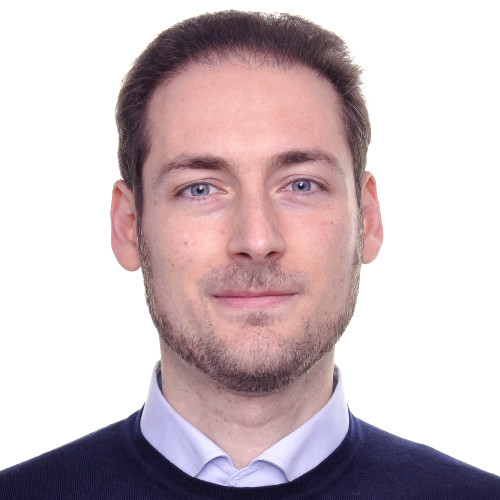}}]{Roberto Natella} (Ph.D.) is Associate Professor at Università degli Studi di Napoli Federico II, Naples, Italy. His research interests are in software security and dependability, with main recurring theme on the experimental injection of faults, attacks, and stressful conditions in software systems. He is co-founder and scientific consultant at Secureware s.r.l.
\end{IEEEbiography}







\end{document}